 \documentclass[prl,amsmath,amssymb,twocolumn, showpacs, superscriptaddress,10pt]{revtex4-1}

\usepackage{amsmath}
\usepackage{hyperref}
\usepackage{graphicx}
\usepackage{amsfonts}
\usepackage{amsthm}
\usepackage{cases}
\usepackage{bm}

\newtheorem{theorem}{Theorem}
\newtheorem{lemma}[theorem]{Lemma}

\usepackage{color}
\definecolor{Blue}{rgb}{0.00, 0.00, 1.00}
\definecolor{Red}{rgb}{1.00, 0.00, 0.00}

\hypersetup{
    colorlinks=true,       % false: boxed links; true: colored links
    linkcolor=red,          % color of internal links (change box color with linkbordercolor)
    citecolor=blue,        % color of links to bibliography
    filecolor=magenta,      % color of file links
    urlcolor=cyan           % color of external links
}

\newcommand{\be}{\begin{equation}}
\newcommand{\ee}{\end{equation}}
\newcommand{\bea}{\begin{eqnarray}}
\newcommand{\eea}{\end{eqnarray}}

\allowdisplaybreaks

\newcommand{\R}{\mathbb{R}}

\usepackage{graphicx}
\newcommand{\ind}{\mathbf{1}}

\newcommand{\Arate}{I_\text{Airy}}
\newcommand{\set}[1]{{\{#1\}}}
\renewcommand{\Im}{\text{Im}}
\renewcommand{\Re}{\text{Re}}
\newcommand{\img}{\mathbf{i}}
\newcommand{\Res}{\text{Res}}
\newcommand{\aden}{\mu_\text{Airy}}
\newcommand{\aDen}{M_\text{Airy}}

\newcommand{\beq}{\begin{equation}}
\newcommand{\eeq}{\end{equation}}
\newcommand{\beqn}{\begin{eqnarray}}
\newcommand{\eeqn}{\end{eqnarray}}
\DeclareMathOperator{\Airy}{Airy}

\begin{document}

%\title{Edge statistics of fermions in a $d$-dimensional hard wall potential}
%\title{Statistics of fermions in a $d$-dimensional box near a hard wall}
\title{Coulomb-gas electrostatics controls large fluctuations of the KPZ equation}

\author{Ivan Corwin}
\affiliation{Columbia University, Department of Mathematics 2990 Broadway, New York, NY 10027 USA}
\author{Promit Ghosal}
\affiliation{Columbia University, Department of Statistics 1255 Amsterdam, New York, NY 10027 USA}
\author{Alexandre Krajenbrink}
\affiliation{CNRS - Laboratoire de Physique Th\'eorique de l'Ecole Normale Sup\'erieure, 24 rue Lhomond, 75231 Paris Cedex, France}
\author{Pierre Le Doussal}
\affiliation{CNRS - Laboratoire de Physique Th\'eorique de l'Ecole Normale Sup\'erieure, 24 rue Lhomond, 75231 Paris Cedex, France}
\author{Li-Cheng Tsai}
\affiliation{Columbia University, Department of Mathematics 2990 Broadway, New York, NY 10027 USA}

\date{\today}

\begin{abstract}
We establish a large deviation principle for the Kardar-Parisi-Zhang (KPZ) equation, providing precise control over the left tail of the height
distribution for narrow wedge initial condition. Our analysis exploits
an exact connection between the KPZ one-point distribution and the Airy point process -- an infinite particle Coulomb-gas which arises at the spectral edge in random matrix theory. We develop the large deviation principle for the Airy point process and use it to compute, in a straight-forward and assumption-free manner, the KPZ large deviation rate function in terms of an electrostatic problem (whose solution we evaluate).
This method also applies to the half-space KPZ equation, showing that its rate function is half of the full-space rate function.
In addition to these long-time estimates, we provide rigorous
proof of finite-time tail bounds on the KPZ distribution which demonstrate a crossover between exponential decay with exponent $3$ (in the shallow left tail) to exponent $5/2$ (in the deep left tail).
The full-space KPZ rate function agrees with the one computed in Sasorov et al. [{ J. Stat. Mech}, 063203 (2017) \cite{sasorov2017large}] via a WKB approximation analysis of a non-local, non-linear integro-differential equation generalizing Painlev\'e II which Amir et al. [Comm. Pure Appl. Math.
{\bf 64}, 466 (2011) \cite{ACQ11}] related to the KPZ one-point distribution.
\end{abstract}


%05.40.-a: Fluctuation phenomena, random processes, noise, and Brownian motion
%02.10.Yn	Matrix theory
%02.50.-r	Probability theory, stochastic processes, and statistics

\maketitle

Since its birth in 1986, the Kardar--Parisi--Zhang (KPZ) equation \cite{KPZ}
%has found applications in numerous domains, from
has been applied to describe growth of interfaces \cite{TS2010}, transport in one-dimension (1D) and
Burgers turbulence \cite{forster1977large}, directed polymers \cite{directedpoly}, chemical reaction fronts \cite{Atis2015}, bacterial growth \cite{bact}, slow combustion \cite{combustion}, coffee stains \cite{yunker2013effects}, conductance fluctuations in Anderson localization \cite{Anderson}, polar active fluids \cite{toner}, Bose Einstein superfluids \cite{Altman}, quantum entanglement growth \cite{nahum2017quantum}.

Whereas some stochastic models (e.g. exclusion processes \cite{johansson}, random permutations \cite{BDJ}, random walks in random media \cite{randommedia}) are directly related (via mappings to `height functions') to the universality class for the 1D KPZ equation; others -- namely random matrix theory (RMT) -- rely on hidden connections to KPZ which are only seen from exact solutions to both KPZ and RMT models \cite{BCMacdonald}. In this Letter, we describe such a relationship between the KPZ equation and the Airy point process -- an infinite particle Coulomb-gas \cite{Serfaty} which arises at the spectral edge in random matrix theory --  and exploit variational techniques of electrostatics to precisely quantify the large fluctuations for the KPZ equation.%; in particular computing its large deviation principle (LDP) rate function.

The 1D KPZ equation describes the stochastic growth of an interface of height $h(t,x)$ at $x \in \mathbb{R}$ and time $t>0$
\be
\label{eq:KPZ}
\partial_t h = \partial_x^2 h + (\partial_x h)^2 + \, \xi(t,x) \;,
\ee
in convenient units, starting from an initial condition $h(t=0,x)$. Here $\xi(x,t)$ is a centered Gaussian white noise with $\overline{\xi(t,x) \xi(t',x')}= 2 \delta(x-x')\delta(t-t')$ and $\overline{\cdots}$ denotes expectations w.r.t. this noise. Typically, the fluctuations of the height field scale, at large time, like $t^{1/3}$. Recent progress has yielded exact solutions for the probability density function (PDF) of the height at a given space point at arbitrary time when starting from special initial conditions (e.g.\ droplet, flat, stationary) \cite{KPZrefs1,ACQ11,KPZrefs2}. Focusing here and below
on the droplet (a.k.a narrow wedge) initial condition, $h(0,x)=- \frac{|x|}{\delta} - \ln(2 \delta)$ for $\delta \ll1$, the
exact formula for the PDF is expressed in terms of a Fredholm determinant. Using this, the scaled and centered height $\mathcal{H}(t)/t^{1/3}$, where $\mathcal{H}(t)=h(t,0)+ \frac{t}{12}$, was shown to converge in law as $t \to +\infty$ to the Tracy-Widom GUE distribution, which also describes the fluctuations of the largest eigenvalue, $\lambda_{\rm max}$, of a large random matrix from the Gaussian unitary ensemble (GUE).

Despite considerable interest, much less is known about large deviations and tails of the KPZ field or PDF $P(H,t)= \frac{\partial}{\partial H} \mathbb{P}(\mathcal{H}(t)\leq H)$.
For general non-equilibrium systems, large deviation rate functions play a role similar to the free energy or entropy in equilibrium systems (see \cite{Nickle} and references therein). Existing large deviation theories fail to apply in the KPZ growth setting. The macroscopic fluctuation theory \cite{MFT} requires local thermodynamic equilibrium, %--- a property which is not present
not realized here. The weak noise theory (see, e.g. \cite{KK1}) applies, but only
at very short times.
%is limited to studying very short times.
Understanding the large deviations for the KPZ equation poses an important conceptual challenge.

Quantitative control over the tails of the KPZ equation plays an important role in experimental and numerical works. Precise results can
%For instance, precise results for the KPZ equation can
be used e.g. as benchmarks for broadly applicable numerical Monte-Carlo methods such as used in \cite{NumericsHartmann}.
In experimental work (such as reviewed in \cite{HH-TakeuchiReview}), the tail behavior we are probing corresponds to excess growth. While unlikely at a single point,
%At a single location along the growing substrate, such a large deviation is quite unlikely. However, seen a whole,
if the growing substrate is sufficiently long, %then
disparate regions (spaced as
%on the order of
time$^{2/3}$) will see roughly independent growth. Hence, by standard extreme-value theory, the maximal and minimal height of the entire %growth
substrate will be determined by the one-point tail behaviors. The KPZ equation also models semiconductor film growth \cite{films}. In technological applications, the roughness of these films determines device performance. As many films are grown independently, large deviations dictate failure rates.
%there may be very
%Many semiconductor films being grown independently, large deviations translate into failure rates.

In population growth and mass transport models, the KPZ tails play contribute to multi-fractal intermittency \cite{Khoshnevisan}. The $H/t^{1/3} \gg 1$ tail is associated with excess mass growth which comes from locally favorable effects; in contrast, the $- H/t^{1/3} \gg 1$ tail is associated with mass die-out which arises from collective effects of wide-spread unfavorable growth regions. Due to this collective effect, the left tail is intrinsically more difficult to analyze at large time. A similar situation arises in RMT for the tails of the PDF of $\lambda_{\rm max}$:  while positive fluctuations arise from the largest eigenvalue $\lambda_{\rm max}$ simply detaching from the bulk of the spectrum, negative ones requires a reorganisation of the entire Wigner semicircle density of eigenvalues (the pushed Coulomb-gas) \cite{dean}. This analogy leads to the prediction \cite{LargeDevUs} that for $t\gg 1$ and large fluctuations $|H|\sim t$ the right tail ($H\gg0$) scales as $- \ln P(H,t) \sim t$ while the left tail ($H\ll0$) scales as $- \ln P(H,t)\sim t^2$.

%To understand better the role of large fluctuations in growth, it is thus important to develop a precise description of the nature of KPZ tails, valid at all times.
%Since methods such as the macroscopic fluctuation theory \cite{Bertini_MFT,bodineau2004current,Derrida_largedeviation} fail for KPZ, new
%Contrary to (simpler) diffusive interacting particle systems, . New approaches are thus required, e.g., as done here, to exploit the aforementioned exact solutions.
For short times $t\ll1$ the left tail of the PDF ($H\ll 0$) behaves as $P(H,t)~\sim~\exp(-\frac{4}{15\pi}|H|^{5/2}/t^{1/2})$, as was shown
analytically (via weak noise theory and exact solutions) \cite{KK1,KMS,LMRS}
and numerically \cite{NumericsHartmann} (see also \cite{MKV,Shorttime2} for other initial conditions). Extracting this tail in the intermediate or
large time limit is much harder. For $t\gg 1$, in the typical scaling region $H \sim t^{1/3}$, the left tail should behave like the Tracy-Widom GUE distribution, i.e.\ $P(H,t) \sim_{H \to -\infty} \exp(- \frac{1}{12} |H|^3/t)$. Until recently, nothing was known about how far this cubic exponent persists into the very far left tail region $|H| \sim t^{\alpha}$ with $\alpha>1/3$, or whether it holds for intermediate times.

%Until recently, nothing is known rigorously about the far tail region for $|H| \sim t^{\alpha}$ with $\alpha>1/3$ at large time, and even less about the tails at intermediate time. In mathematics, the results obtained so far
%%this problem has been considered in several works \cite{Mueller1991,Mueller2008,MorenoFlores,Khoshnevisan} however all of these results
%only give upper bounds on the tail decay (with the wrong power on $|H|$ in the exponential) and are
%not adapted to the scaling by $t^{1/3}$ and centering required in the large time limit.
%%also
%%not adapted to the centering by $t/12$ and scaling by $t^{1/3}$
%%which one must take as time goes to infinity.

Given the similarities between the KPZ and RMT problems, it is natural to try to attack these tail questions using methods inspired by RMT.
The left tail behavior for $\lambda_{\rm max}$ can be accessed by either (i) the Coulomb-gas and associated electrostatic variational problem for the GUE spectrum \cite{GuionnetGUE,dean} (see also \cite{MSreview} for other large deviation applications of the Coulomb gas) or (ii) the relationship between gap probabilities and certain classical integrable systems \cite{AdlerVan} (which, in $N\to \infty$ edge limit, relate to the Painlev\'{e} II equation \cite{TWAll}).
\cite{ACQ11} introduced a non-local, non-linear integro-differential equation which generalizes Painlev\'{e} II by including a ``Fermi-factor", and showed that its solution relates to the KPZ PDF. Studying this generalized equation via standard `integrable-integral operator' methods \cite{IIKS} involves infinite-dimensional Riemann-Hilbert problem steepest descent analysis which is beyond current techniques. Employing a certain approximation ansatz, \cite{LargeDevUs} attempted to analyze this equation. While they successfully predicted the scaling form for the large deviation tail $P(H,t) \sim \exp(- t^2 \Phi_{-}(H/t))$  for $-H \sim t \gg1$,  the approximations were too reductive and \cite{LargeDevUs} predicted $\Phi_{-}(z) = \frac{1}{12}|z|^3$ which turns out only to hold true for $z$ near 0. Ref. \cite{sasorov2017large} revisited this analysis and employed a WKB approximation along with a `self-consistency' ansatz for the form of the solution to a Schr\"odinger equation in which the potential depends upon the solution. Given these assumptions, \cite{sasorov2017large} extracted a formula
\begin{equation}\label{eq:Phi}
\Phi_-(z)=\frac{4}{15\pi^6}(1-\pi^2 z)^{5/2}-\frac{4}{15\pi^6}+\frac{2}{3\pi^4}z -\frac{1}{2\pi^2}z^2
\end{equation}
which predicts a crossover between $\Phi_{-}(z) \simeq_{z \to -\infty} \frac{4}{15 \pi}  |z|^{5/2}$ and $\simeq_{z \to -0}\frac{1}{12} |z|^{3}$. This, taken with the short-time estimates, suggests that the $|H|^{5/2}$ tail remains valid at all times (see also \cite{NumericsHartmann} and \cite{KK1,MKV,KMS})
and that there is a crossover between the $\frac{1}{12}|H|^3/t$ and $\frac{4}{15\pi}|H|^{5/2}/t^{1/2}$ tail when $|H|\approx t$ (once $t\gg 1$).

\smallskip
The purpose of this Letter is to demonstrate how the Coulomb-gas can be utilized in a straight-forward and assumption-free manner to (i) establish, using the large deviations for the Airy point process, an electrostatic variational formula for $\Phi_{-}(z)$ whose solution (which we derive) agrees with \eqref{eq:Phi}, and (ii) demonstrate the first precise tail bounds \eqref{eq:1} which are valid for all intermediate and long times and which capture the crossover between the $\frac{1}{12}|H|^3/t$ tail for $|H|\ll t$ and the $\frac{4}{15 \pi}|H|^{5/2}/t^{1/2}$ tail for $|H|\gg t$. Our work provides a description of the intermediate and late time left large deviations for the KPZ equation where the connection to RMT and the role of the collective effects is explicit: each fixed value of $z=H/t$ corresponds to an optimal eigenvalue density (see Fig. \ref{Fig:den}). Finally, we extend our study to the half-line KPZ equation in the critical case, which relates to the Gaussian orthogonal ensemble (GOE), leading (via our RMT approach) to the rate function $\Phi^{\text{half-space}}_{-}(z) = \frac{1}{2} \Phi^{\text{full-space}}_{-}(z)$.

Our starting point is a remarkable identity \cite{borodin2016moments, B16}, obtained from the exact solution of the droplet initial condition \cite{KPZrefs1,ACQ11} which directly connects KPZ and RMT (as well as fermions in an harmonic well at temperature of order $t^{-1/3}$ \cite{dean2015finite}): for $\varphi_{t,s}(a)=\log(1+e^{t^{\frac{1}{3}}(a+s)})$
\begin{equation}
\overline{\exp(-e^{\mathcal{H}(t) + s t^{1/3}})} =\mathbb{E}_{\Airy}\Big[\exp\Big(-\sum_{i=1}^\infty \varphi_{t,s}(\mathbf{a}_i )\Big)\Big].\label{id}
\end{equation}
The l.h.s is an expectation over the KPZ white noise giving access to $P(H,t)$ while the r.h.s is the expectation of a ``Fermi factor" over the {\it Airy point process} (Airy PP) generating the set $\lbrace \mathbf{a}_i\rbrace \in \mathbb{R}$. The Airy PP describes the largest few eigenvalues of a large GUE matrix. It is a `determinantal' measure on infinite point configurations $\mathbf{a}=(\mathbf{a}_1>\mathbf{a}_2>\cdots)$ on $\mathbb{R}$ which means that for all $k\geq 1$, the $k$-th correlation function $\rho_k(x_1,\ldots,x_k)$ (which equals the probability density for the event that $\big\{x_i\in \mathbf{a}, \textrm{ for all } 1\leq i\leq k\big\}$) takes the form $\rho_k(x_1,\ldots,x_k) = \det\big(K(x_i,x_j)\big)_{1\leq i,j\leq k}$ for some fixed `correlation kernel' $K:\mathbb{R}^2\to \mathbb{C}$. The Airy PP correlation kernel is $K_{\mathbf{Ai}}(x,y) =  \int_{0}^{\infty} \mathbf{Ai}(x+r) \mathbf{Ai}(y+r)\mathrm{d}r$. In particular the mean density is $\rho(a)=\rho_1(a)=K_{\mathbf{Ai}}(a,a) \simeq_{a \to -\infty} \pi^{-1} \sqrt{|a|}$. This agrees with the square-root behavior of the Wigner semi-circle at the edge. Remarkably, the $|H|^{5/2}$ tail emerges quite simply from this $\sqrt{|a|}$ density as we show from the first term in the cumulant expansion of the r.h.s.\ of \eqref{id}, see \eqref{k1}. After observing this, we describe the Airy PP large deviation principle (LDP) derived via Coulomb-gas, and use it to compute the full crossover rate function $\Phi_-(z)$. Finally, we provide the bounds \eqref{eq:1} which describes intermediate time behavior of the tail.

%Before proving Theorem \ref{thm:main}, we show how a cumulant expansion  provides the $|H|^{5/2}$ behavior, and how through a simple conditioning of the APP, we can derive a reasonable upper-bound on $\Phi_{-}(z)$ which fairly closely agrees with  \eqref{eq:Phi} (especially for $z\to 0$ and $z\to -\infty$).

\smallskip
\noindent{\bf Cumulant expansion.}
As $st^{\frac{1}{3}}\to \infty$ the l.h.s.\ of \eqref{id}
approaches $\mathbb{P}(\mathcal{H}(t) \leq -s t^{\frac{1}{3}})=\mathbb{P}(\mathcal{H}(t) \leq zt)$ with $z=-st^{-2/3}$. The r.h.s.\ of \eqref{id} is evaluated via cumulants as
\begin{equation} \label{cum1}
\log\big( \textrm{r.h.s.(\ref{id})}\big)=\sum_{n=1}^\infty \frac{\kappa_n}{n!}
\end{equation}
where $\kappa_n$ %is the term of order $\varphi_{t,s}^n$ in the expansion and
is the $n$-th cumulant of the Airy PP whose general form is known  \cite{Soshnikov},
e.g.\ for $n=1,2$
\bea
\kappa_1 = -\mathrm{Tr}(\varphi_{t,s} K_{\mathbf{Ai}}) = -\int_{-\infty}^{+\infty} \mathrm{d}a \, \varphi_{t,s}(a) \rho(a).
\eea
and $\kappa_2=\mathrm{Tr}(\varphi_{t,s}^2K_{\mathbf{Ai}})-\mathrm{Tr}(\varphi_{t,s} K_{\mathbf{Ai}} \varphi_{t,s} K_{\mathbf{Ai}})$, where $(\psi K)(x,y)=\psi(x) K(x,y)$, $\mathrm{Tr} K = \int_{\mathbb{R}} \mathrm{d}a\,  K(a,a)$. In the limit $z  \to - \infty$, it is sufficient to keep only the first cumulant (the $n=1$ term) in \eqref{cum1}, which, using the above asymptotics
$\rho(a) \simeq_{a \to -\infty} \pi^{-1} \sqrt{|a|}$, is estimated as (we use the notation $(\cdot)_{+}=\max(\cdot,0)$ below)
\bea
 \kappa_1 &\simeq& -t^{1/3} \int_{-\infty}^{+\infty} \mathrm{d}a (a+s)_+ \rho(a) \nonumber \\
& \simeq& -t^{1/3} \frac{4}{15\pi}s^{5/2} = - t^2\frac{4}{15\pi}|z|^{5/2}.  \label{k1}
\eea
This simple argument
gives the leading behavior as $z\to-\infty$ of the left large deviation rate function, $\Phi_-(z) \simeq \frac{4}{15\pi} |z|^{5/2}$, hence the desired $|H|^{5/2}$ tail. % (it also provides an upper bound for all $z$).
Explicit calculation (see \cite{KrajLedou2018}) of the next higher cumulants
\begin{equation}
\kappa_2 \simeq t^{2/3}\frac{s^2}{\pi^2 } = t^2 \frac{z^2}{\pi^2 }
 \; , \; \kappa_3 \simeq
- t \frac{4|s|^{3/2}}{\pi^3} =- t^2 \frac{4|z|^{3/2}}{\pi^3} \label{k2}
\end{equation}
shows their subdominance both (i) for $-z\gg 1$ with $t\gg1$ and $z=H/t$ fixed and (ii) $t$ fixed and large $s=-H/t^{1/3}$ and reproduces the large $|z|$ expansion of \eqref{eq:Phi}. %We checked that %the two next cumulants
%\eqref{k2} reproduces exactly the next two terms in the large $|z|$ expansion of \eqref{eq:Phi}. %Interestingly, for  \eqref{eq:Phi} to hold, all even order cumulants, starting from $\kappa_8$, must be 0  \cite{KrajLedou2018}. %We now obtain the full function $\Phi_{-}(z)$ by a variational method.

% in the expansion of the prediction of \cite{sasorov2017large} in
%Interestingly, for this matching to work, all even order cumulants starting at $\kappa_8$ must vanish {\red Is this the correct statement; Reword if not}.

\smallskip
\noindent{\bf Coulomb-gas and large deviation rate function.} Using \eqref{id}, $\Phi_{-}(z)$ can be computed as (write $\mathbb{E}$ for $\mathbb{E}_{\Airy}$)
$$
\Phi_{-}(z)=\lim_{t\to \infty} \frac{1}{t^2} \log \mathbb{E}\Big[\exp\Big(-\sum_{i=1}^\infty \varphi_{t,-zt^{2/3}}(\mathbf{a}_i )\Big)\Big].
$$
For large $t$, we have $\varphi_{t,-zt^{2/3}}(t^{2/3}a)\approx t(a-z)_+$.
Let $\mu_{t}(a)\mathrm{d}a = t^{-1} \sum_{i\geq 1} \delta_{-t^{-2/3}\mathbf{a}_i}(a)\mathrm{d}a$ denote the \emph{scaled}, space-reversed Airy PP empirical measure.
Then we have
\begin{align}
\label{eq:Phi:}
\begin{split}
	\Phi_{-}(z)=
	\lim_{t\to \infty} &\frac{1}{t^2} \log \mathbb{E}\Big[
\exp\Big(-t^2 \int_{\mathbb{R}}\mathrm{d}a \,\mu_t(a)(-z-a)_+ \Big)\Big].
\end{split}
\end{align}
%The APP should (though it has not been proven) enjoy a large deviation principle so that for a suitable class of function $\mu$,
Like the GUE, the Airy PP should enjoy an LDP so that for a suitable class of functions $ \mu $,
$\mathbb{P}(\mu_t\approx \mu)\approx \exp\big(-t^{2} I_\text{Airy}(\mu)\big)$. To our knowledge, this rate function is not in the literature, and we describe it below and in \cite{SuppMat}.
%for a rate functional $I(\cdot)$. Assuming this, we finally find that
Given this, the r.h.s.\ of~\eqref{eq:Phi:} can be evaluated
via a variational problem, $ \Phi_{-}(z)=\min_{\mu} \Sigma(\mu) $,
with cost function
\begin{align}
	\label{eq:kpzvar}
	\Sigma(\mu) =
	\int_{\mathbb{R}}\mathrm{d}a \, \mu(a)(-z-a)_{+} +I_\text{Airy}(\mu).
\end{align}
%where the minimum is over the class of functions upon which $I$ is finite. We know that $I$ is minimized (and equal to zero) for $\mu$ given by the limit of the APP density (i.e., $\mu_{*}(a) = \pi^{-1} \sqrt{-a}\mathbf{1}_{a\leq 0}$). Plugging in this choice we recover the computation made earlier via cumulants.

To derive the LDP for the Airy PP we will appeal to the fact that the Airy PP arises as an edge limit of the GUE. The GUE spectrum is a 1D Coulomb-gas with logarithmic interaction which immediately leads to an electrostatic variational formulation for the GUE LDP \cite{Serfaty,GuionnetGUE} (with the Wigner semi-circle representing the minimizer of this electrostatic energy). Our approach is to rewrite the GUE LDP in such a manner that it admits an edge scaling limit to yield the Airy PP LDP.

Recall from~\cite{GuionnetGUE} that the empirical measure $ \Lambda_N(\lambda)\mathrm{d}\lambda = \frac{1}{N} \sum_{i=1}^N \delta_{\boldsymbol{\lambda}_i}(\lambda)\mathrm{d}\lambda $
associated to the eigenvalues $\{\boldsymbol{\lambda}_1,\ldots,\boldsymbol{\lambda}_N\}$ of the GUE (normalized to have typical support $[-2,2]$ -- see \cite{SuppMat} for a precise definition)
%{\blue LC: In the supplementary material, I put $ \boldsymbol\xi_i $ as the pre-scale GUE and $ \boldsymbol\lambda_i=N^{-1/2}\boldsymbol\xi_i $. So here I change the definition of $ \Lambda_N $, previously it was $\Lambda_N(\lambda)\mathrm{d}\lambda = \frac{1}{N} \sum_{i=1}^N \delta_{N^{-1/2}\boldsymbol{\lambda}_i}(\lambda)\mathrm{d}\lambda $.}
%{\red I: The unscaled GUE measure should be specified precisely in the supp mat, with a reference to that here}
enjoys an LDP so that, for a generic density $ \Lambda $ with unit mass,
$ \mathbb{P}(\Lambda_N \approx \Lambda)\approx \exp\big(-N^{2} I_2(\Lambda)\big) $. The rate function $I_{\beta}(\Lambda)$ is the difference of the electrostatic energy of a Coulomb-gas of charge $\beta$ (with $\beta=2$ for GUE, and $\beta=1$ for GOE) with density $\Lambda$, as compared to that of the Wigner semi-circle density $ \Lambda_\text{sc}(\lambda) = \frac{1}{2\pi}\sqrt{4-\lambda^2}\ind_\set{|\lambda|<2} $.
%
% (we write this for general Gaussian $\beta$-ensembles, noting that $\beta=2$ is GUE and $\beta=1$ is GOE) $I_\beta(\Lambda) = \beta (E(\Lambda)-E(\Lambda_\text{sc}))$
%is the difference between the electrostatic energy $ E(\Lambda) $ {\red I: this needs to be defined somewhere or at least references in the supp mat}
%%$
%%	E(\Lambda) = \frac{1}{4}\int \mathrm{d}\lambda \, \lambda^2 \Lambda(\lambda)
%%	- \frac{1}{2} \int\log|\lambda_1-\lambda_2|\prod_{i=1}^2 \mathrm{d}a_i\Lambda_i(\lambda_i)
%%$
%compared to that of
%%the equilibrium (minimizer) density $ \Lambda_\text{eq} = \text{argmin}_{\Lambda'} E(\Lambda') $.
%%The latter is given by  Wigner's
%the semicircle law
%%$ \Lambda_\text{eq}(\lambda) =
%$ \Lambda_\text{sc}(\lambda) = \frac{1}{2\pi}\sqrt{4-\lambda^2}\ind_\set{|\lambda|<2} $.
$I_{\beta}$ can be rewritten (see~\cite{SuppMat} for details) as
\begin{align}
	\label{eq:Ibeta}
	I_\beta(\Lambda)
	=
	\frac{\beta}{2} J(\Lambda)
	+
	\frac{\beta}{2} \int_{\R} \mathrm{d}\lambda \, V(\lambda) \Lambda(\lambda),
\end{align}
with a Coulomb interaction term
$
	J(\Lambda)
	=
	-\int_{\R^2} \log|\lambda_1-\lambda_2|
	\prod_{i=1}^2 \mathrm{d} \lambda_i (\Lambda(\lambda_i)-\Lambda_\text{sc}(\lambda_i))
$ (note that $\Lambda-\Lambda_\text{sc}$ is a signed density with integral over $\R$ equal to 0)
and potential term
$ V(\lambda) = \int_0^{|\lambda|} \mathrm{d}\lambda' \, (({\lambda'}^2-4)_+)^{1/2} $.
%which solves the stationary equation
%\begin{align}
%	\label{eq:sc}
%	\frac{\lambda^2}{4} - \int_{\R} \mathrm{d}\lambda' \log|\lambda-\lambda'| \Lambda_\text{sc}(\lambda') = \text{const.}
%\end{align}
%Using ~\eqref{eq:sc} we rewrite $ I_\beta(\Lambda) $ in a compact form as
%\begin{align}
%	\label{eq:Ibeta}
%%	I_\beta(\Lambda) =
%	- \frac{\beta}{2} \int_{\R^2} \log|\lambda_1-\lambda_2|
%	\prod_{i=1}^2 \mathrm{d} \lambda_i \big(\Lambda(\lambda_i)-\Lambda_\text{sc}(\lambda_i)\big),
%\end{align}
%in terms of centered (signed) density $ (\Lambda-\Lambda_\text{sc}) $ that has zero mass.
%
The (space-reversed) Airy PP %$ \{-\mathbf{a_1}<-\mathbf{a_2}<\ldots\} $
arises as a scaling limit %$ \{N^{\frac13}(\boldsymbol{\lambda}_i-2)\} $
of the GUE spectrum near its lower edge $ \lambda=-2 $.
To deduce the Airy PP LDP from that of the GUE,
we introduce the scaling $ \lambda=-2+t^{2/3}N^{-2/3}a $.
%with $ t\gg1 $ large but fixed.
%
%Recall that we have defined $\mu_t$ the empirical measure of the sAPP,
%$\mu_{t}(b)\mathrm{d}b = t^{-1} \sum_{i\geq 1} \delta_{-t^{-2/3}\mathbf{a}_i}(b)\mathrm{d}b$.
%observe that
%{\red I: This argument is confusing. Why don't you use Pierre's notation from the end of the paper? Using $\lambda$ and $a$ to represent both scaled and unscaled quantities is confusing.}
As $ N \to \infty $,
%the number of points of scaled-GUE in a given interval $[\lambda,\lambda+\mathrm{d}\lambda]$
%should match that of the scaled-APP in the corresponding interval $ [a,a+\mathrm{d}a ] $.
%Schematically matching
$ N \mathrm{d}\lambda\, \Lambda_N(\lambda) \simeq t \mathrm{d} a \, \mu_t(a) $, %, and replacing
which when inserted into \eqref{eq:Ibeta} %$ N \mathrm{d}\lambda \Lambda(\lambda) = t \mathrm{d}b \mu(b) $
gives $ N^2 I_\beta(\Lambda) \simeq t^2 I_\text{Airy}(\mu) $,
with
$$I_\text{Airy}(\mu) = J_\text{Airy}(\mu) + U(\mu). $$
%\begin{align*}
%	I_\text{Airy}(\mu) = - \int \log|b_1-b_2| \prod_{i=1}^2 \mathrm{d}b_i \big(\mu(b_i)-\mu_\text{Airy}(b_i)\big),
%\end{align*}
%
%{\red Ivan: I find this argument difficult to accept. The sc measure certainly has square-root edge behavior; but why does $\Lambda$ need to?}
%Extracting square-root scaling factors $ \sqrt{t^{2/3}N^{-2/3}} $ from the densities
%(as observed from $ \Lambda_\text{sc}(\lambda) $ near $ \lambda=-2 $),
%we gain an overall factor of $ t^{2}N^{-2} $,
%%\begin{align*}
%%	\mathbb{P}(\mu_t\approx \mu) \approx \lim_{N\to\infty} \big\{ \exp(-N^2 I_2(\mu_{t,N})) \big\}.
%%\end{align*}
%%Evaluating the r.h.s.\ assuming $ \lim_N\mu_{t,N}=\mu_t $,
%which in turns gives an LPD of
%APP at speed $ t^2 $ (see Supp.\ Mat.\ for details):
%$ \mathbb{P}_\text{Airy}(\mu_t\approx \mu) \approx \exp(-t^2 I_\text{Airy}(\mu)) $, with
%the rate function
Here ${J_\text{Airy}(\mu)\!=\!-\!\int\!\log|a_1\!-\!a_2|\!\prod_{i=1}^2\!\mathrm{d}a_i(\mu(a_i)\!-\!\mu_\text{Airy}(a_i))}$
is defined for densities $ \mu $ satisfying mass-conservation $ \int \mathrm{d}a\, (\mu(a)-\mu_\text{Airy}(a))=0 $,
where $ \mu_\text{Airy}(a) = \frac{1}{\pi}\sqrt{a} \mathbf{1}_{\{a>0\}}$,
and $ U(\mu) = \frac{4}{3}\int_{-\infty}^0 \mathrm{d}a\, |a|^{\frac32} \mu(a) $.

%To find the minimum of~\eqref{eq:kpzvar},
Instead of searching directly for the minimum of $ \Sigma $ in~\eqref{eq:kpzvar},
we first consider a simpler cost function
$$
	\Sigma_J(\mu)
	=
	\int_{\mathbb{R}}\mathrm{d}a \, (-z-a)_{+}\mu(a) +J_\text{Airy}(\mu)
$$
that drops the term $ U(\mu) $.
The minimizer $\mu_*$ of $\Sigma_J$ is the unique measure (see \cite{SuppMat} for details) such that
\begin{align}\label{eq:SigmaJmin}
(-z-a)_+ - 2\int_{\R} \mathrm{d}a' \log|a-a'|(\mu_*(a')-\mu_\text{Airy}(a'))\geq \text{c}
\end{align}
for some constant $\text{c}$ with strict equality on the support of $\mu_*$.
Differentiating the l.h.s. of \eqref{eq:SigmaJmin} in $a$ yields
\begin{align}\label{eq:SigmaJmindiff}
-\mathbf{1}_{\{a<-z\}} - 2\int_{\R} \mathrm{d}a' \frac{\mu_*(a')-\mu_\text{Airy}(a')}{a-a'}.
\end{align}
Consider a generic interval $[u,\infty)$ and let
\begin{align*}
	\mu_{*,u}(a)
	=
	\Big(
		&\frac{1}{\pi}\sqrt{a-u} + \frac{1}{2\pi^2}
		\log\Big|\frac{\sqrt{a-u}+\sqrt{v}}{\sqrt{a-u}-\sqrt{v}}\Big|
\\
	&+\frac{1}{\pi}\Big(\frac{u}{2}-\frac{\sqrt{v}}{\pi}\Big)\frac{1}{\sqrt{a-u}}\Big)\ind_\set{a>u},
\end{align*}
where $ v= -z-u $. Ref. \cite{SuppMat} verifies that substituting this density $\mu_{*,u}(a)$ for $\mu_{*}(a)$ implies that \eqref{eq:SigmaJmindiff}$=0$ on $[u,\infty)$. Furthermore, \cite{SuppMat} shows that $u= u_0 = \frac{2}{\pi^2}(\sqrt{1-\pi^2z}-1) $ is the unique choice of $u$ for which for which one also has \eqref{eq:SigmaJmindiff}$\geq 0$ on $(-\infty,u_0)$ and $=0$ on $[u_0,\infty)$. This means that $\mu_*(a)=\mu_{*,u_0}(a)$ satisfies \eqref{eq:SigmaJmin} and hence is the unique minimizer of $\Sigma_J$.  Evaluating yields (see  Fig~\ref{Fig:den})
\begin{align*}
	\mu_{*}(a)
	=
	\Big(\frac{1}{\pi}\sqrt{a-u_0}+ \frac{1}{2\pi^2} \log\Big|\frac{\sqrt{a-u_0}+\frac{\pi}{2}u_0}{\sqrt{a-u_0}-\frac{\pi}{2}u_0}\Big|\Big)\ind_\set{a>u_0}.
\end{align*}
The associated minimum of $\Sigma_J$ is
\begin{align*}
	\min_{\mu} \Sigma_J(\mu) %%&= \Sigma_J(\mu_{*,u_0})
	&=\frac{4}{15\pi^6}(1-\pi^2 z)^{\frac52}-\frac{4}{15\pi^6}+\frac{2}{3\pi^4}z -\frac{1}{2\pi^2}z^2,
\end{align*}
which coincides precisely with $\Phi_{-}(z)$ in \eqref{eq:Phi}.

Returning to $ \Sigma $ from \eqref{eq:kpzvar}, we note that $ U(\mu)\geq 0 $ implies $ (\min\Sigma) \geq (\min \Sigma_J) $.
Since  $ \mu_*(a)$ vanishes for $ a<0 $ (since $ u_0>0 $), we have $ U(\mu_*)=0 $ and hence $ \Sigma(\mu_*)=\Sigma_J(\mu) $.
Thus, the minimizer and minimum for $ \Sigma_J $ in fact also applies to $ \Sigma $. Since $ \Phi_{-}(z)=\min_{\mu} \Sigma(\mu) $, this confirms the formula in \eqref{eq:Phi} and the calculation of \cite{sasorov2017large}.

\begin{figure}[ht]
\includegraphics[width=.9\linewidth]{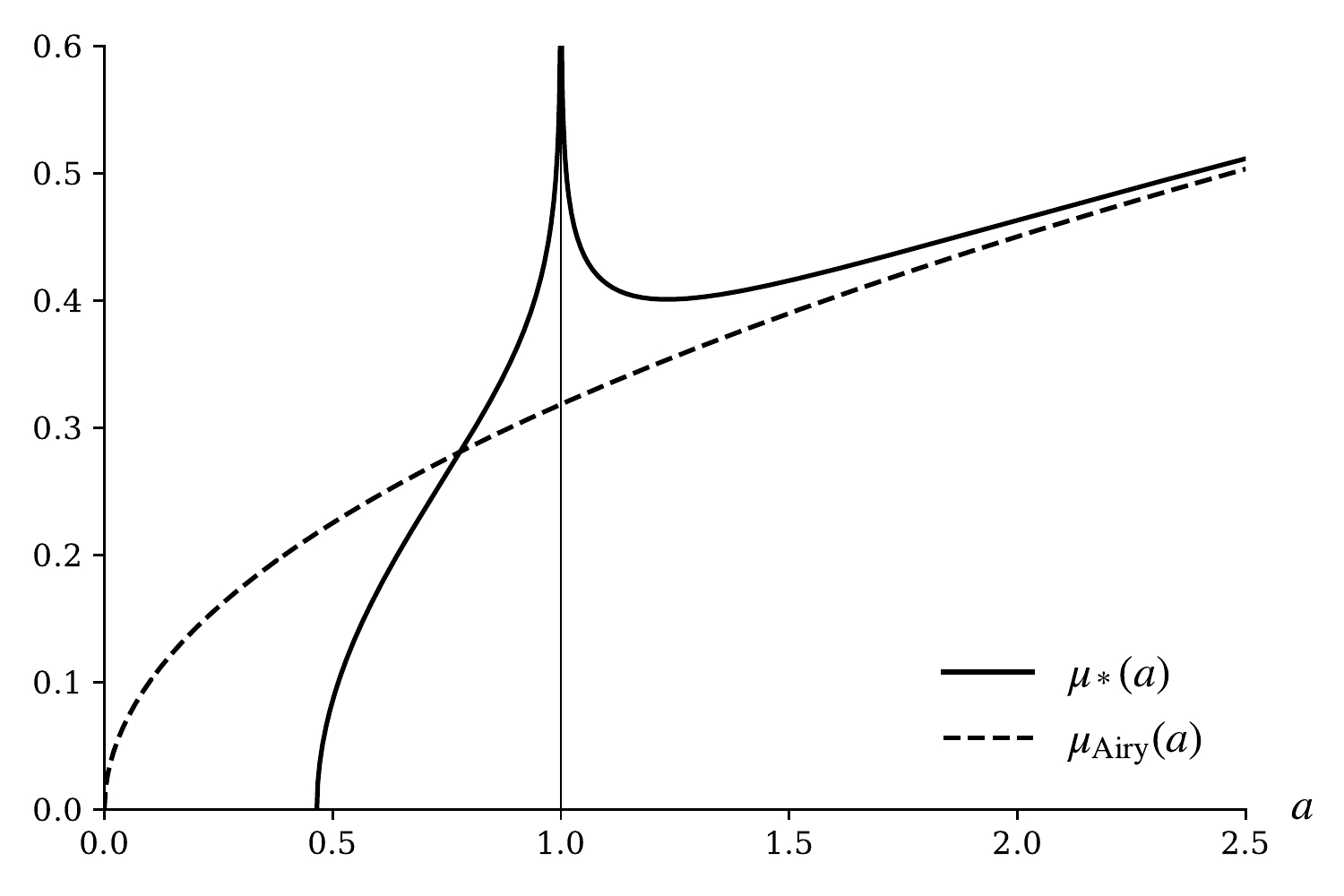}
\caption{Optimal density $ \mu_{*}(a)$ at $ z=-1 $ compared to  $\mu_{\Airy}(a)$. The density $ \mu_* $ has a log singularity at $ a=-z $ (c.f. \cite{SatyaGardschTaxier} for another Coulomb gas problem with similar behavior).}
\label{Fig:den}
\end{figure}

%\smallskip
\noindent {\bf Tail bounds for intermediate times.} While the KPZ LDP holds for $t\to \infty$, the crossover behavior between exponents $3$ and $5/2$ remains valid at all intermediate times. Precisely: For any $\varepsilon, \delta\in ]0,\frac{1}{3}[$ and $t_0>0$ then there exists constants $S=S(\varepsilon,\delta,t_0)$, $K_1=K_1(\varepsilon,\delta,t_0)>0$ and $K_2=K_2(t_0)>0$ such that for all $s\geq  S$ and $t\geq t_0$,
\begin{align}\label{eq:1}
\nonumber\mathbb{P}(H\! \leq \! -st^{\frac{1}{3}})\! &\leq \! e^{-\frac{4(1-\varepsilon)}{15 \pi}t^{\frac{1}{3}}s^{\frac{5}{2}}}+e^{-K_1 s^{3-\delta}-\varepsilon t^{\frac{1}{3}}s}+e^{-\frac{1-\varepsilon}{12}s^3}\\
\mathbb{P}(H\! \leq \! -st^\frac{1}{3})\! &\geq e^{-\frac{4(1+\varepsilon)}{15\pi}t^{\frac{1}{3}}s^{5/2}}+e^{-K_2 s^3}
\end{align}
%\begin{equation}\label{eq:1}
%\mathbb{P}(H\! \leq \! -st^{\frac{1}{3}})\! \leq \! e^{-\frac{4(1-\varepsilon)}{15 \pi}t^{\frac{1}{3}}s^{\frac{5}{2}}}+e^{-K_1 s^{3-\delta}-\varepsilon t^{\frac{1}{3}}s}+e^{-\frac{1-\varepsilon}{12}s^3}
%\end{equation}
%%and
%\begin{equation}\label{eq:2}
%\mathbb{P}(H\! \leq \! -st^\frac{1}{3})\! \geq e^{-\frac{4(1+\varepsilon)}{15\pi}t^{\frac{1}{3}}s^{5/2}}+e^{-K_2 s^3}
%\end{equation}

For $t^{2/3} \gg s \gg 1$, the second and third terms in the first line of \eqref{eq:1} dwarf the first term and represent cubic decay (in the exponential) in $s$. In particular, as $t$ gets large, only the third term survives and we recover (up to an $\varepsilon$ correction) the predicted $\frac{1}{12}s^3$ decay. On the other hand, for $s\gg t^{2/3}$ the first term in the second line of  \eqref{eq:1} dwarfs the others and recovers the predicted $\frac{4}{15\pi}s^{5/2}$ decay for all $t$. The second line of \eqref{eq:1} contains corresponding lower bounds -- though notice that for $t$ large and $t^{2/3}\gg s\gg 1$, our bounds do not recover the $\frac{1}{12}$ constant for the lower bound on the cubic decay. This result recovers the large and small $z$ behavior of the $t\to \infty$ rate function $\Phi_{-}(z)$. Prior to \eqref{eq:1}, the only finite time bounds were in \cite{Mueller1991etc} which provided a Gaussian upper-bound on the decay (hence, the wrong exponent). Moreover, those bounds are not adapted to large $t$ center and scaling---becoming ineffective as $t$ grows.
%
%A suitably strong large deviation principle (LDP) for the APP should enable a proof of a formula for $\Phi_{-}(z)$. No such LDP exists, though there is relevant work for the related GUE \cite{GuionnetGUE,dean2006large,dean2008extreme,majumdar2014top,PerretSchehr}.

Eqs. \eqref{eq:1} follow from two considerations. The typical locations of the $\mathbf{a}_i$ are governed by $\rho(a)$. Plugging these typical values into \eqref{id} yields the $5/2$ exponential term. However, the $\mathbf{a}_i$ are random and may deviate from their typical locations. For instance $\mathbf{a}_1\leq -s$ with probability  $\approx \exp(-\frac{1}{12}s^3)$. Such deviations lead to the cubic exponential terms. In order to provide matching upper and lower tail bounds, we precisely control the LDP for the counting function of the Airy PP in large intervals. This can be done via asymptotics of the Ablowitz-Segur solution to Painlev\'{e} II \cite{AS70s,Bothner15} which relates to the exponential moment generating function for this counting process, as well as by using of the relation of the AAP to the stochastic Airy operator \cite{StochasticAiry}. The main ideas and steps of this derivation are provided in \cite{SuppMat} (and further technical details and complete rigorous proofs are in \cite{CorwinGhosalTail}).

\smallskip
\noindent{\bf Extensions and Summary.} The approach developed in this Letter is applicable to certain variants of the KPZ equation which enjoy identities similar to \eqref{id} -- namely half-space KPZ \cite{barraquand2017stochastic}, the stochastic six vertex model and ASEP \cite{B16,BO16}. Briefly we consider the half-space KPZ equation, i.e. \eqref{eq:KPZ} restricted to $x \in \mathbb{R}^+$, with Neumann b.c. $\partial_x h(t,x)|_{x=0}=A$, for the value $A=-1/2$ corresponding to the so-called critical case. In that case and for droplet initial condition \cite{barraquand2017stochastic} proved that
$$
\overline{\exp(-\frac{1}{4}e^{\mathcal{H}(t) + s t^{1/3}})} = \mathbb{E}_{\mathrm{GOE}}\Big[\prod_{i=1}^{+\infty} \frac{1}{\sqrt{1+e^{t^{1/3} (\mathbf{a}_i+s)}}} \Big]
$$
where the r.h.s.\ expectation is over the $\beta=1$ version of the Airy PP (which describes the top few eigenvalues at the spectral edge for the GOE instead of GUE -- see also \cite{SuppMat}). Employing the Airy PP Coulomb-gas approach from this Letter, we find that due to the square-root in the r.h.s. above (which introduces a factor of $1/2$ in exponential form), and the value of $\beta=1$ (instead of $\beta=2$), the half-space KPZ rate function $\Phi^{\text{half-space}}_{-}(z) = \frac{1}{2} \Phi_{-}(z)$ where $\Phi_{-}(z)$ is the full-space function in \eqref{eq:Phi}. More generally, for the partition sum defined in Eq. (1.12) in Ref. \cite{VadimSodin18} for general $\beta$, we obtain $\Phi^{\beta}_{-}(z)
= \frac{\beta}{2} \Phi_{-}(z)$ (see \cite{SuppMat} for details).
%There are variants of the KPZ / APP identity which hold true for other choices of initial data such as half-Brownian or stationary, for which our approach should readily produce LDP rate functions.
 Finally, there are good reasons to conjecture that for (full space stationary) Brownian IC
$\Phi^{\rm Br}_{-}(z)=\Phi^{\rm droplet}_{-}(z)$, see  \cite{SuppMat}.

In conclusion, by relating the distribution of the height for the KPZ equation to an expectation over the Airy point process, we are able to employ the Coulomb-gas formalism and associated electrostatic problem large deviation principle (first for the GUE and, through a limit transition which we present, for the Airy point process) to identify the KPZ rate function. Solving the variational problem produces the formula in \eqref{eq:Phi}. This argument brings the role of random matrix theory in the study of KPZ to the forefront and provides a straight-forward and assumption-free derivation of the KPZ rate function. Additionally, a similar approach should be applicable to other exactly solvable KPZ class models such as ASEP or the stochastic six vertex model which connect to discrete Coulomb-gases. This approach also permits us to derive results valid for all intermediate times and opens the way to systematically calculate higher order corrections between the long time and finite time PDF, as is useful in experiments and numerics.

\begin{acknowledgments}
{\it Acknowledgments:}
We thank G. Barraquand,
A. Borodin, T. Bothner, E. Corwin, P. Deift, T. Halpin-Healy, S. N. Majumdar, B. Meerson, G. Schehr, H. Spohn and B. Virag for helpful discussions and comments. The authors initiated this work during the 2017 session of PCMI, partly funded by the NSF grand DMS:1441467. I.C. was partially funded by the NSF grant DMS:1664650 and the Packard Foundation through a Packard Fellowship for Science and Engineering. PLD and AK acknowledge support from ANR grant ANR-17-CE30-0027-01 RaMaTraF. LCT was partially supported by a Junior Fellow award from the Simons Foundation, and by the NSF through DMS:1712575.
\end{acknowledgments}

%{\red Why are some titles present in the bibliography and others not present? Shouldn't we be consistent?}
{}

\newpage
\mbox{}

\newpage

%%%%%%%%%%%%%%%% supplementary material section if we need it %%%%%%%%%%%%

\appendix

\newpage

\begin{widetext}

\bigskip

\bigskip

\begin{large}
\begin{center}

Supplementary Material for {\it Coulomb-gas electrostatics controls large fluctuations of the KPZ equation}

\end{center}
\end{large}

%
%\author{Ivan Corwin}
%\affiliation{Columbia University, Department of Mathematics 2990 Broadway, New York, NY 10027 USA}
%\author{Promit Ghosal}
%\affiliation{Columbia University, Department of Statistics 1255 Amsterdam, New York, NY 10027 USA}
%\author{Alexandre Krajenbrink}
%\affiliation{CNRS - Laboratoire de Physique Th\'eorique de l'Ecole Normale Sup\'erieure, 24 rue Lhomond, 75231 Paris Cedex, France}
%\author{Pierre Le Doussal}
%\affiliation{CNRS - Laboratoire de Physique Th\'eorique de l'Ecole Normale Sup\'erieure, 24 rue Lhomond, 75231 Paris Cedex, France}
%\author{Li-Cheng Tsai}
%\affiliation{Columbia University, Department of Mathematics 2990 Broadway, New York, NY 10027 USA}
%
%
%
%\date{\today}
%
%\begin{abstract}
We give the principal details of some of the calculations described in the main text of the Letter.
%\end{abstract}

%\pacs{05.40.-a, 02.10.Yn, 02.50.-r}

%05.40.-a: Fluctuation phenomena, random processes, noise, and Brownian motion
%02.10.Yn	Matrix theory
%02.50.-r	Probability theory, stochastic processes, and statistics

\maketitle

\appendix

\section{A) Deriving the rate functions}
%{\red I: You should recall the LDP statement for the GUE involving $E$ so that the derivation below makes sense}
\label{SecGUErt}

We will consider the GUE spectrum, which can be seen as the $\beta=2$ charge case of a Coulomb-gas of $N$ particles in a harmonic potential defined by
$\mathbb{P}\big(\{\boldsymbol{\lambda}_1< \cdots< \boldsymbol{\lambda}_N\}\big) =  \frac{1}{Z_N}\exp(-\beta  (\frac{N}{4}\sum_{i=1}^N\lambda_i^2-\frac12\sum_{i\neq j}\log|\lambda_i-\lambda_j|))$ (here $Z_{N}$ is the normalization needed to make this a probability measure).
Under this choice of scaling, the spectrum tends to be supported in $[-2,2]$.
The empirical measure $ \Lambda_N(\lambda) := N^{-1} \sum_{i=1}^N \delta_{\boldsymbol{\lambda}_i}(\lambda) $ converges as $N\to \infty$ to the Wigner semi-circle distribution with density $\Lambda_\text{sc}(\lambda) = \frac{1}{2\pi}\sqrt{4-\lambda^2}\ind_\set{|\lambda|<2}$.
Recall from~\cite{GuionnetGUE} that $\Lambda_N(\lambda)$ enjoys an LDP, so that for a given probability density function $ \Lambda $,
$ \mathbb{P}(\Lambda_N\approx\Lambda) \approx \exp(-N^2 I_\text{2}(\Lambda)) $, where
$ I_\beta(\Lambda) = \beta (E(\Lambda)-E(\Lambda_\text{sc})) $,
and
\begin{align*}
%	\label{eq:eEnergy}
	E(\Lambda) = \frac{1}{4}\int \mathrm{d}\lambda \, \lambda^2 \Lambda(\lambda)
	- \frac{1}{2} \int\log|\lambda_1-\lambda_2|\prod_{i=1}^2 \mathrm{d}\lambda_i\Lambda(\lambda_i).
\end{align*}
We explain how to rewrite this rate function as \eqref{eq:Ibeta}.
First, the semicircle law $ \Lambda_\text{sc} $, being the minimizer of $ E $ among the set of all probability distributions, satisfies
\begin{align}
	\label{eq:sc:}
	\frac{\lambda^2}{4} - \int_{\R} \mathrm{d}\lambda' \log|\lambda-\lambda'|\Lambda_\text{sc}(\lambda')
	=
	\frac12 V(\lambda)
	+\text{const.}
\end{align}
for some potential $ V $ that vanishes  $ V|_{\lambda\in[-2,2]}=0 $ within the support $ [-2,2] $ of $ \Lambda_\text{sc} $,
and is nonnegative $ V|_{|\lambda|>2} \geq 0 $ off the support.
%
%where $ V(\lambda) = \int_0^{|\lambda|} \mathrm{d}\lambda' \, \sqrt{({\lambda'}^2-4)_+} $.
%The function $ V $ vanishes for $ |\lambda|<2 $,
%and \eqref{eq:sc:} gives the stationary condition $ (\delta E/\delta \Lambda)(\Lambda_\text{sc}) =0 $
%that $ \Lambda_\text{sc} $ satisfies as a minimizing density of $ E $,
%where $ \Lambda $ varies among all densities with unit mass that has support within the support $ [-2,2] $ of $ \Lambda_\text{sc} $.
%{\red I: what is the "stationary condition" and do week need to specify support $[-2,2]$?}
More explicitly, by calculating $ \int_{-2}^2 \frac{\mathrm{d}\lambda'\,\Lambda_\text{sc}(\lambda')}{\lambda-\lambda'} $,
and then integrating the result in $ \lambda $, we find
\begin{align*}
	V(\lambda) = \int_0^{|\lambda|} \mathrm{d}\lambda' \, \sqrt{({\lambda'}^2-4)_+} =\Big( \frac{|\lambda|}{2} \sqrt{\lambda^2-4} - 2\log\big(\sqrt{\lambda^2-4}+|\lambda|\big) +2\log(2)\Big)\ind_\set{|\lambda|>2}
\end{align*}
with $V(\lambda) \simeq \frac{4}{3} \big(|\lambda|-2\big)^{3/2} \ind_\set{|\lambda|>2}$ for $|\lambda|\approx 2$.
Now, given a generic density $ \Lambda $ with unit mass, we write $ \Lambda = (\Lambda-\Lambda_\text{sc})+\Lambda_\text{sc} $,
and insert this into the electrostatic energy function $ E(\Lambda) $
%{\red I: This formula for $E$ should have occurred displayed earlier}
%$
%	E(\Lambda) = \frac{1}{4}\int \mathrm{d}\lambda \, \lambda^2 \Lambda(\lambda)
%	- \frac{1}{2} \int\log|\lambda_1-\lambda_2|\prod_{i=1}^2 \mathrm{d}a_i\Lambda_i(\lambda_i)
%$
to get
\begin{align*}
	E(\Lambda) - E(\Lambda_\text{sc})
	=
	\frac12 J(\Lambda)
	+
	\int_{\R} \mathrm{d}\lambda
	\big(\Lambda(\lambda)-\Lambda_\text{sc}(\lambda)\big)	
	\bigg(		
		\frac{{\lambda}^2}{4} - \int_{\R} \mathrm{d}\lambda' \log|\lambda-\lambda'| \Lambda_\text{sc}(\lambda')
	\bigg),
\end{align*}
where
$
	J(\Lambda) = -\int_{\R^2} \log|\lambda_1-\lambda_2|
	\prod_{i=1}^2 \mathrm{d} \lambda_i (\Lambda(\lambda_i)-\Lambda_\text{sc}(\lambda_i)).
$
We may substitute \eqref{eq:sc:} into the last term above.
Recall that we normalized the empirical measure $ \Lambda_N $ so as to have total mass $1$, which implies $ \int_{\R}\mathrm{d}\lambda (\Lambda(\lambda)-\Lambda_\text{sc}(\lambda))=0 $. Thus, the constant in~\eqref{eq:sc:} does not contribute after integrating over $ \lambda $.
From these considerations we obtain
\begin{align}
	\label{eq:Ibeta:}
	I_\beta(\Lambda)
	=
	\beta ( E(\Lambda) - E(\Lambda_\text{sc}) )
	=
	\frac{\beta}{2} J(\Lambda)
	+
	\frac{\beta}{2} \int_{\R} \mathrm{d}\lambda \, V(\lambda) \Lambda(\lambda) = \frac{\beta}{2} I_2(\Lambda).
\end{align}
Note that in deriving this, we have also used the fact that the r.h.s.\ of~\eqref{eq:Ibeta:} indeed vanishes for $ \Lambda=\Lambda_\text{sc} $ (since the potential $ V(\lambda)=0 $ on the support $ [-2,2] $ of $ \Lambda_\text{sc} $).
Hence, we have derived the claimed formula \eqref{eq:Ibeta}.

We find it more convenient (in accordance with earlier work of \cite{dean}) to work with the space-reversed Airy PP which arises as a scaling limit of the GUE spectrum near its lower edge $ \lambda=-2 $.
To relate the GUE LDP to the Airy PP LDP,
we introduce the scaling $ \lambda=-2+t^{2/3}N^{-2/3}a $.
In the $ N \to \infty $ limit, $ N \mathrm{d}\lambda\, \Lambda_N(\lambda) \simeq t \mathrm{d} a \, \mu_t(a) $
in \eqref{eq:Ibeta:}, which gives %$ N^2 I_\beta(\Lambda) \simeq t^2 I_\text{Airy}(\mu) $,
\begin{align*}
	N^2 I_2(\Lambda)
	\simeq
	&-
	t^2 \int_{\R^2} \log|a_1-a_2| \prod_{i=1}^2 \mathrm{d} a_i (\mu(a_i)-t^{-1/3}N^{1/3}\Lambda_\text{sc}(-2+t^{2/3}N^{-2/3}a_i))
\\
	&-
	t^2 \int_{\R} \mathrm{d} a \mu(a) t^{-1}N V(-2+t^{2/3}N^{-2/3}a_i).
\end{align*}
Taking $ N\to\infty $ on the r.h.s., with $ t $ fixed, we obtain
the Airy PP LPD rate function $ I_\text{Airy}(\mu) = J_\text{Airy}(\mu) + U(\mu) $, where
\begin{align*}
	J_\text{Airy}(\mu) =- \int \log|a_1-a_2| \prod_{i=1}^2 \mathrm{d}a_i \big(\mu(a_i)-\mu_\text{Airy}(a_i)\big),
	\quad
	U(\mu) = \frac{4}{3}\int_{-\infty}^0 \mathrm{d}a\, |a|^{\frac32} \mu(a),\quad \aden(a)=\frac{1}{\pi}\sqrt{a}\ind_\set{a>0},
\end{align*}
%and $ \aden(a)=\frac{1}{\pi}\sqrt{a}\ind_\set{a>0} $.
and where we assume that all candidate densities $\mu$ satisfy mass-conservation $ \int \mathrm{d}a\, (\mu(a)-\mu_\text{Airy}(a))=0 $.
%holds at the GUE level because there are always exactly $N$ points.
%A priori, such a condition could be lost upon passing to the edge limit. However, as it turns out, the AAP \emph{does} preserve the zero mass condition.
%To see this, we note $ J_\text{Airy}(\mu) $ can be expressed, via integration by parts, as
%\begin{align*}
%	J_\text{Airy}(\mu)
%	=
%	\frac{1}{2} \int_{\R^2} \mathrm{d}a\mathrm{d}a' \Big(\frac{\int_a^{a'}\mathrm{d}b\,(\mu(b)-\aden(b))}{a'-a}\Big)^2
%	+
%	\lim_{(a,a')\to(-\infty,\infty)}
%	\Big\{ \Big(\int_a^{a'}\mathrm{d}b\,(\mu(b)-\aden(b))\Big)^2\log|a'-a|\Big\}.
%\end{align*}
%The last term is infinite for those density $ \mu $ with $ \int_{-\infty}^\infty \mathrm{d}b\,(\mu(b)-\aden(b)) \neq 0 $.
%
%As a by-product, this calculation also confirms that $ I_\text{Airy}(\mu) $ is nonnegative,
%which must hold for a rate function.

\section{B) Details on solving the electrostatic variational problem}\label{SecVarDet}

Here we calculate
%Now, set $ \beta=2 $, we wish to find
the minimum and minimizer of
\begin{align*}
	\Sigma(\mu) := A \int_0^{a_0} \mathrm{d}a\, (a_0-a)\mu(a) + \Arate(\mu),
\end{align*}
for given $ A,a_0 \geq 0 $. The relevant case is $ A=1 $ and $ a_0=-z $.
We keep the dependence on $ A $ to demonstrate how $ \Sigma $ scales with $ A $.

Instead of solving this problem directly, as noted earlier, we consider first a simpler cost function
\begin{align*}
	\Sigma_J(\mu) := A \int_0^{a_0} \mathrm{d}a\, (a_0-a)\mu(a) + J_\text{Airy}(\mu)
	=
	A \int_0^{a_0} \mathrm{d}a\, (a_0-a)\mu(a) - \int_{\R^2} \log|a_1-a_2| \prod_{i=1}^2 \mathrm{d}a_i (\mu(a_i)-\aden(a_i)).
\end{align*}
%To solve this variational problem, we postulate that the minimizer $ \mu_* $ has density $ \mu_*(a) $ and has a connected support $ [u,\infty) $ (where $u$ is a parameter we will determine later).
%Looking at $ \delta \Sigma(\mu)/\delta\mu $ among all measures $ \mu $ supported in $ [u,\infty) $ such that $ \int_\R \mathrm{d} a(\mu(a)-\aden(a)) =0 $, we find that
For such variational problem,
a measure $ \mu_* $ with support $ [u_0,\infty) $ is the unique minimizer if
\begin{subequations}
\label{eq:var}
\begin{align}
	&A(a_0-a)_+ - 2\int_{\R} \mathrm{d}a' \log|a-a'| (\mu_{*}(a')-\mu_\text{Airy}(a')) =c,
	\quad\text{for } a>u_0,
\\
	&A(a_0-a)_+ - 2\int_{\R} \mathrm{d}a' \log|a-a'| (\mu_{*}(a')-\mu_\text{Airy}(a')) \geq c,
	\quad \text{for } a<u_0,
\end{align}
\end{subequations}
where $ c $ denotes a constant.
This  criterion holds for the analogous variational problem for GUE (see, e.g.\ \cite[Theorem~2.6.1, Lemma~2.6.2]{AGZ}) and goes through to the Airy PP limit.
%Taking an edge limit indicates that ~\eqref{eq:var} is a sufficient condition for $ \mu_* $ being the unique minimizer.
%where $ \mu_\text{Airy}(a) = \frac{1}{\pi}\sqrt{a}\mathbf{1}_{\{a>0\}} $.
Differentiating shows that for \eqref{eq:var} to hold, it suffices that
\begin{subequations}
\label{eq:inteq}
\begin{align}
	\label{eq:inteq1}
	&-A\ind_\set{a<a_0} - 2\int_{\R} \mathrm{d} a' \, \frac{\mu_{*}(a')-\mu_\text{Airy}(a')}{a-a'} =0,
	\quad
	\text{for }a>u_0,
\\
	\label{eq:inteq2}
	&	-A\ind_\set{a<a_0} - 2\int_{\R} \mathrm{d} a' \, \frac{\mu_{*}(a')-\mu_\text{Airy}(a')}{a-a'} < 0,
	\quad \text{for } a<u_0.
\end{align}
\end{subequations}
The integral in~\eqref{eq:inteq} is \emph{principal value} and the same holds for subsequent integrals even when not explicitly stated.

In \eqref{eq:var}, both the minimizing density $ \mu_{*} $ and $ u_0 $ are unknown.
Our strategy of solving this problem is to first consider a \emph{generic} $ u $ in place of $ u_0 $,
and solve the integral equation
\begin{align}
	%\tag{\ref{eq:inteq1}'}
	\label{eq:inteq:}
	-A\ind_\set{a<a_0} - 2\int_{\R} \mathrm{d} a' \, \frac{\mu_{*,u}(a')-\mu_\text{Airy}(a')}{a-a'} =0,
	\quad
	a>u,
\end{align}
for a density $ u_{*,u} $ supported in $ [u,\infty) $.
Later, we will identify a unique $ u_0 $ such that \eqref{eq:inteq2} holds for $ u_{*,u_0}:=u_* $.

\subsection{B.1) Solving the integral equation~\eqref{eq:inteq:}.}
Set $ v=a_0-u $. We claim that the solution to ~\eqref{eq:inteq:} is
\begin{align}
	\label{eq:mu*u}
	\mu_{*,u}(a)
	:=
	\Big(
		\frac{1}{\pi}(a-u)^{\frac12} + \frac{1}{\pi}\Big(\frac{u}{2} - \frac{Av^{\frac12}}{\pi}\Big) (a-u)^{-\frac12}
		+ \frac{A}{2\pi^2} \log \Big|\frac{\sqrt{a-u}+\sqrt{v}}{\sqrt{a-u}-\sqrt{v}}\Big|
	\Big)\ind_\set{a>u}.
\end{align}
Such a solution can be obtained from the taking edge limit of the solution to the analogous equation for the GUE, solved in a way similar to \cite{dean}.
For what is relevant here, we will just \emph{verify} that $ \mu_{*,u} $ does solve~\eqref{eq:inteq:}.

It is convenient to decompose $\mu_{*,u}-\mu_\text{Airy} = \mu_1+\mu_2 $,
where
\begin{align*}
	\mu_1(a)
	&:=
	\Big( \frac{1}{\pi}(a-u)^{\frac12} + \frac{u}{2\pi}(a-u)^{-\frac12} \Big)\ind_\set{a>u}	
	-
	\frac{1}{\pi}a^{\frac12} \ind_\set{a>0},
\\
	\mu_2(a)
	&:=
	\frac{A}{\pi^2}
	\Big(
		-v^{\frac12}(a-u)^{-\frac12}
		+ \frac{1}{2} \log \Big|\frac{\sqrt{a-u}+\sqrt{v}}{\sqrt{a-u}-\sqrt{v}}\Big|
	\Big)\ind_\set{a>u}.
\end{align*}

In the following, we calculate $ f_i(a) = - \int_{\R}\mathrm{d}a' \frac{\mu_i(a')}{a-a'} $.
For the purpose of verifying that $ \mu_{*,u} $ solves~\eqref{eq:inteq:}, we need to know $ f_i(a) $ for $ a>u $.
Ultimately, we will also need to know for $ a<u $ to identify $ u_0 $ through \eqref{eq:inteq2}.
We begin with $ f_1 $.
\begin{align*}
	f_1(a)
	= -\int_{\R} \mathrm{d} a'\, \frac{\mu_1(a')}{a-a'}
	&= \frac{1}{\pi} \int_u^\infty \mathrm{d} a'\, \Big( \frac{(a'-u)^{\frac12}}{a'-a} + \frac{u}{2}\frac{(a'-u)^{-\frac12}}{a'-a} \Big)
	   - \frac{1}{\pi} \int_0^\infty \mathrm{d} a' \, \frac{{a'}^{\frac12}}{a'-a}.
%\\
%	&= \frac{1}{\pi} \int_0^\infty \Big( \frac{y}{y^2-x} + \frac{u}{2}\frac{y^{-1}}{y^2-x} - \frac{y}{y^2-(x+u)} \Big)2y dy
%\\
%	&= \frac{1}{\pi} \int_{\R} \Big( \frac{y^2}{y^2-x} + \frac{u}{2}\frac{1}{y^2-x} - \frac{y^2}{y^2-(x+u)} \Big) dy.
\end{align*}
Performing the change of variables $ (a'-u)^{\frac12} \mapsto w $ in the first integral and $ {a'}^{1/2}\mapsto w $ in the second integral yields
\begin{align*}
	f_1(a)
	&= \frac{1}{\pi} \int_0^\infty \mathrm{d}w \, 2w \Big( \frac{w}{w^2-(a-u)} + \frac{u}{2}\frac{w^{-1}}{w^2-(a-u)} - \frac{w}{w^2-a} \Big).
\end{align*}
The integrand is now an even function of $ w $. Symmetrizing, we extend it integral to the full line $ \R $ and find that
\begin{align*}
	f_1(a)
	=
	\int_{\R} \mathrm{d} w \tilde{f}_1(w) ,
	\quad
	\tilde{f}_1(w)
	=
	\frac{1}{\pi}
	\Big( \frac{{w}^2}{{w}^2-(a-u)} + \frac{u}{2}\frac{1}{{w}^2-(a-u)} - \frac{{w}^2}{{w}^2-a} \Big).
\end{align*}
We evaluate this integral separately for the cases $ a>u $, $ 0<a<u $, and $ a<0 $ via complex analysis.
As noted earlier, only case $ a>u $ is relevant toward verifying $ \mu_{*,u} $ solves~\eqref{eq:inteq:},
and the cases $ 0<a<u $ and $a<0$ are in place for a later purpose.\\

\noindent \underline{The case $ a>u $}:
%To this end, we check that the integrand decays as $ O(|z|^{-4}) $ for large $ |z| $.
The integrand has four poles at $ w=\pm\sqrt{a-u}, \pm\sqrt{a} $ along the real.
This being the case, integrating $ \tilde{f}_1(w) $ along the contour $ \Gamma(R,\delta;\sqrt{a-u},\sqrt{a}) $ as depicted in Fig.~\ref{fig:cntur} yields zero.
Further, it is readily checked that the integrand $ \tilde{f}_1(w) $ decays as $ O(|w|^{-4}) $ as $ |w|\to\infty $.
Given these properties, letting $ R\to\infty $ and $ \delta\to 0 $ gives
%\begin{align*}
%	f_1(x) = \frac{1}{\pi}\sum_{a=\pm \sqrt{x}, \pm\sqrt{x+u}} \int_{\gamma_\delta(a)}
%	\Big( \frac{z^2}{z^2-x} + \frac{u}{2}\frac{1}{z^2-x} - \frac{z^2}{z^2-(x+u)} \Big) dz,
%\end{align*}
%where $ \gamma_\delta(a) := \{|z-a|=\delta : \Im(z)>0\} $ is a contourclockwise oriented upper half circle,
%centered at $ a $ with small radius $ \delta $.
%Letting $ \delta\downarrow 0 $ gives
\begin{align*}
	f_1(a) =
	\sum_{w_0=\pm \sqrt{a-u}, \pm\sqrt{a}}
	\pi\img \,
	\underset{w=w_0}{\Res}
	\big[ \tilde{f}_1(w) \big] =0.
%	\frac{\img\pi}{\pi}
%	\Big(
%		\sum_{a=\pm \sqrt{x}} \Big(\frac{a^2}{2a}+\frac{u/2}{2a}\Big)
%		-
%		\sum_{a=\pm\sqrt{x+u}}\frac{a^2}{2a}
%	\Big)
%	=0.
\end{align*}

\noindent\underline{The case $ 0<a<u $}:
In this case the integrand has poles at $ z=\pm \img\sqrt{|a-u|}, \pm\sqrt{a} $.
This being the case, integrating over the contour $ \widetilde{\Gamma}(R,\delta;\sqrt{a}) $
as depicted in Fig.~\ref{fig:cntur} gives $ 2\pi\img\,{\Res}_{w=\img\sqrt{|a-u|}} [ \tilde{f}_1(w) ] $.
Letting $ R\to\infty $ and $ \delta\to 0 $ gives
\begin{align*}
	f_1(a) =
	2\pi\img \,
	\underset{w=\img\sqrt{|a-u|}}{\Res} \big[ \tilde{f}_1(w) \big]
	+
	\sum_{w_0=\pm\sqrt{a}}
	\pi\img \,
	\underset{w=w_0}{\Res} \big[ \tilde{f}_1(w) \big].
%
%	f_1(x)
%	&=
%	\frac{2\pi\img}{\pi} \underset{z=\img\sqrt{|x|}}{\Res}
%		\Big( \frac{z^2}{z^2-x} + \frac{u}{2}\frac{1}{z^2-x}  \Big)
%	+
%	\lim_{\delta\downarrow 0}
%	\frac{1}{\pi}\sum_{\pm\sqrt{x+u}} \int_{a=\gamma_\delta(a)} \Big( - \frac{z^2}{z^2-(x+u)} \Big) dy.
\end{align*}
The last term is zero. Evaluating the first residue gives
\begin{align*}
	f_1(a) = \tfrac{u}{2}|a-u|^{-\frac12} - |a-u|^{\frac12}.
\end{align*}

\noindent\underline{The case $ a<0 $}:
In this case the integrand has poles at $ z=\pm \img\sqrt{|a-u|} $, so
\begin{align*}
	f_1(a) =
	2\pi\img \,
	\underset{w=\img\sqrt{|a-u|}}{\Res} \big[ \tilde{f}_1(w) \big]
	=
	\tfrac{u}{2}|a-u|^{-\frac12} - |a-u|^{\frac12}.
%
%	f_1(x)
%	&=
%	\frac{2\pi\img}{\pi} \underset{z=\img\sqrt{|x|}}{\Res}
%		\Big( \frac{z^2}{z^2-x} + \frac{u}{2}\frac{1}{z^2-x}  \Big)
%	+
%	\lim_{\delta\downarrow 0}
%	\frac{1}{\pi}\sum_{\pm\sqrt{x+u}} \int_{a=\gamma_\delta(a)} \Big( - \frac{z^2}{z^2-(x+u)} \Big) dy.
\end{align*}

\begin{figure}[h]
\includegraphics[width=.95\linewidth]{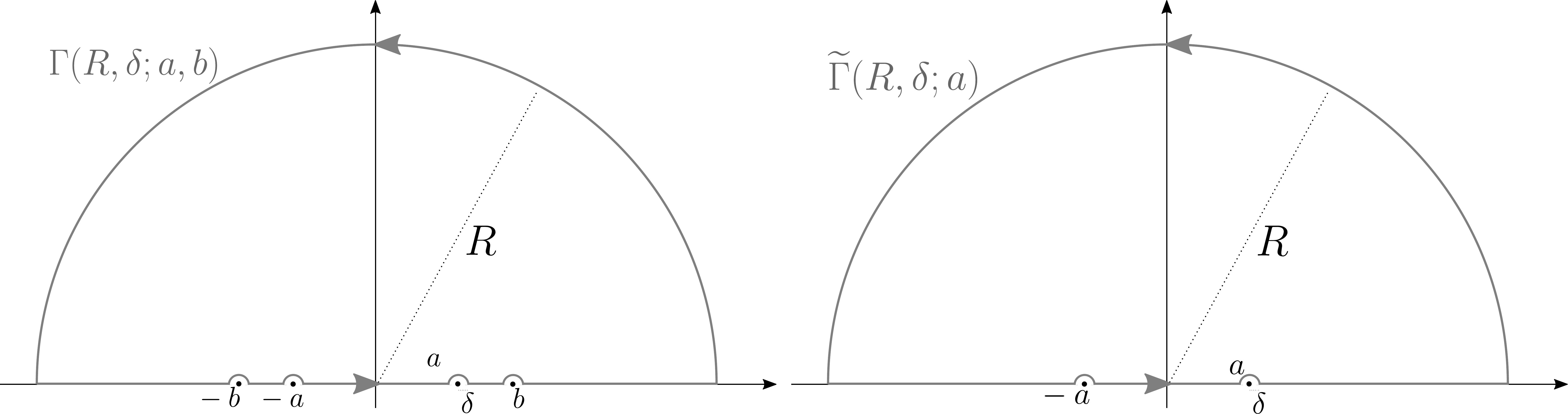}
\caption{The contours $ \Gamma(R,\delta;a,b) $ and $ \widetilde{\Gamma}(R,\delta;a) $}
\label{fig:cntur}
\end{figure}

Next we turn to $ f_2 $.
\begin{align*}
	f_2(a)
	= -\int_{\R} \mathrm{d}a'\frac{\mu_2(a')}{a-a'}
	&=
	\frac{A}{\pi^2}
	\int_{u}^\infty
	\frac{1}{a'-a}	
	\Big(
		-v^{\frac12} (a'-u)^{-\frac12}
		+ \frac{1}{2} \log \Big|\frac{\sqrt{a'-u}+\sqrt{v}}{\sqrt{a'-u}-\sqrt{v}}\Big|
	\Big).
%\\
%	&
%	=
%	\frac{A}{\pi^2}
%	\int_0^\infty
%	\frac{1}{y-x/v}
%	\Big(
%		-y^{-1/2}
%		+ \frac{1}{2} \log \Big|\frac{\sqrt{y}+1}{\sqrt{y}-1}\Big|
%	\Big) dy
%\\
%	&
%	=
%	\frac{A}{\pi^2}
%	\int_{\R}
%	\frac{1}{y^2-x/v}
%	\Big(
%		-1
%		+ \frac{y}{2} \log \Big|\frac{y+1}{y-1}\Big|
%	\Big) dy.
\end{align*}
Performing change of variables $ (a'-u)^{\frac12} \mapsto y $,
followed by symmetrization of the integrals gives
\begin{align*}
	f_2(a)
	= \int_{\R} \mathrm{d} y
	\frac{A}{\pi^2}
	\frac{1}{y^2-(a-u)v^{-1}}
	\Big(
		-1
		+ \frac{y}{2} \log \Big|\frac{y+1}{y-1}\Big|
	\Big).
\end{align*}
To prepare for the complex integrals in the following,
let us consider the integral $ \int_{\R} \mathrm{d} w \tilde{f}_2(w) $
of the function
\begin{align*}
	\tilde{f}_2(w)
	=
	\frac{A}{\pi^2}
	\frac{1}{w^2-(a-u)v^{-1}}
	\Big(
		-1
		+ \frac{w}{2} \log \frac{w+1}{w-1}
	\Big)	
\end{align*}
that is analytic in the upper half plan $ \{\Im[w]>0\} $.
Along the real line $ \R $,
it is readily checked that the imaginary part $ \Im[\tilde{f}_2(y)] $ is an odd function of $ y\in\R $,
and the real part gives exactly the relevant integrand for $ f_2(a) $:
\begin{align*}
	\Re[\tilde{f}_2(y)]
	=
	\frac{1}{y^2-(a-u)v^{-1}}
	\Big(
		-1
		+ \frac{y}{2} \log \Big|\frac{y+1}{y-1}\Big|
	\Big),
	\quad
	y\in\R.
\end{align*}
Hence, $ \tilde{f}_2(a) = \int_\R dw \tilde{f}_2(w) $.
We evaluate this integral separately for the cases $ a>a_0 $, $ u<a<a_0 $, and $ a<u $.\\

\noindent\underline{The case $  a>a_0 $}:
The integrand $ \tilde{f}_2(w) $ is analytic in the upper half plan $ \{\Im(w)>0\} $,
has poles at $ w=\pm\sqrt{(a-u)/v} $ along the real axis, and absolutely integrable singularities at $ w=\pm 1 $.
This being the case, integrating along $ \Gamma(R,\delta,1,\sqrt{(a-u)/v}) $ gives zero.
Further, it is readily checked that $ \tilde{f}_2(w) = O(|w|^{-4}) $ as $ |w|\to\infty $.
Letting $ R\to\infty $ and $ \delta\to 0 $ gives
\begin{align*}
	f_2(a)
	=
	\pi\img \sum_{w_0=\pm\sqrt{(a-u)/v}}
	\underset{w=w_0}{\Res}
	\big[ \tilde{f}_2(w) \big]
%	&= \frac{A}{\pi^2}
%	\Re\Bigg( \sum_{a=\pm\sqrt{x/v}} \int_{\gamma_\delta(a)} \frac{1}{z^2-x/v}\Big(-1+\frac{z}{2} \log \Big(\frac{z+1}{z-1}\Big)\Big) dz \Bigg)
%\\
	&=
	\frac{A\img}{\pi} \sum_{w_0=\pm\sqrt{(a-u)/v}}
	\frac{1}{2w_0}\Big(-1+\frac{w_0}{2} \log \Big(\frac{w_0+1}{w_0-1}\Big)\Big)
\end{align*}
With $ a>a_0=u+v $, we have $ |w_0|=\sqrt{(a-u)/v}>1 $, whereby $ \log (\frac{w_0+1}{w_0-1}) = \log|\frac{w_0+1}{w_0-1}| $. Hence
\begin{align*}
	f_2(a)
	&=
	\frac{A\img}{\pi} \sum_{w_0=\pm\sqrt{(a-u)/v}}
	\frac{1}{2w_0}\Big(-1+\frac{w_0}{2} \log \Big|\frac{w_0+1}{w_0-1}\Big|\Big)
	=0.
\end{align*}

\noindent\underline{The case $ u<a<a_0 $}:
In this case the integrand $ \tilde{f}_2(w) $ has poles at $ w_0=\pm\sqrt{|a-u|/v} $ along the real.
The only difference between the previous case is that, with $ a<a_0 $,
we have $ |w_0|=\sqrt{(a-u)/v}<1 $, whereby
%
%\begin{align*}
%	f_2(x)
%	&=
%	\frac{A}{\pi^2}
%	\Re\Bigg(
%		\img\pi \sum_{a=\pm\sqrt{x/v}} \frac{1}{2a}\Big(-1+\frac{a}{2} \log \Big(\frac{a+1}{a-1}\Big)\Big)
%	\Bigg).
%\end{align*}
%The difference here is that, with $ 0<x<v $, we have
$ \log (\frac{w_0+1}{w_0-1}) = \log|\frac{w_0+1}{w_0-1}| - \img\pi $.
This gives
\begin{align*}
	f_2(a)
	&=
	\frac{A\img}{\pi} \sum_{z_0=\pm\sqrt{(a-u)/v}}
	\frac{1}{2w_0}\Big(-1+\frac{w_0}{2} \log \Big|\frac{w_0+1}{w_0-1}\Big| - \frac{w_0\img\pi}{2}\Big)
	=
	\frac{A}{2}.	
%	
%	&=
%	\frac{A}{\pi^2}
%	\Re\Bigg(
%		\img\pi \sum_{a=\pm\sqrt{x/v}} \frac{1}{2a}\Big(-1+\frac{a}{2} \log \Big|\frac{a+1}{a-1}\Big| - \frac{a\img\pi}{2} \Big)
%	\Bigg)
%	=
%	\frac{A}{2}.
\end{align*}

\noindent\underline{The case $ a<u $}:
In this case the integrand $ \tilde{f}_2(w) $ has poles at $ w_0=\pm\img\sqrt{|a-u|/v} $.
Integrating $ \tilde{f}_2(w) $ along the contour $ \widetilde{\Gamma}(R,\delta;1) $ gives
$ 2\pi\img\,\text{Res}_{w_0=\sqrt{(a-u)/v}}[\tilde{f}_2(w)] $.
Hence, letting $ R\to\infty $ and $ \delta\to 0 $ we obtain
\begin{align*}
	f_2(a)
	&=
	2\pi\img
	\underset{w=\img\sqrt{|a-u|/v}}{\Res}
	\big[ \tilde{f}_2(w) \big]
%\\	
%	\Re\Bigg[
%		\frac{2A\img}{\pi}
%		\underset{z=\img\sqrt{|a-u|v^{-1}}}{\Res}
%		\Big[
%			\frac{1}{z^2+|a-u|v^{-1}}\Big(-1+\frac{z}{2} \log \Big(\frac{z+1}{z-1}\Big) \Big)
%		\Big]
%	\Bigg]
%\\
%	&=
%	\frac{A}{\pi^2}
%	\Re\Big(
%		\frac{2\pi\img}{2\img a}
%		\Big(-1+\frac{\img a}{2} \img\big(2\tan^{-1}(a)-\pi \big)\Big)\Big|_{a=\sqrt{|x|/v}}
%	\Big)
%\\
	=
	\frac{A}{\pi}\Big( -v^{\frac12}|a-u|^{-\frac12} -\tan^{-1}\big(\sqrt{|a-u|/v}\big) + \frac{\pi}{2}  \Big).
\end{align*}

To summarize, we have
\begin{align}\label{faeq}
	f(a) &:=
	-\int_{\R} \mathrm{d} a' \frac{\mu_{*,u}(a')-\aden(a')}{a-a'}
	=
	\frac{A}{2}\ind_\set{a<a_0} + g(a), \quad a\in\R,
\\
	\label{eq:g}
	&
	\text{where }
	g(a)
	=
	\big( - |a-u|^{\frac12}+\big( \tfrac{u}{2} - \tfrac{1}{\pi}Av^{\frac12} \big) |a-u|^{-\frac12} - \tfrac{A}{\pi}\tan^{-1}(\sqrt{|a-u|/v}) \big)
	\ind_\set{a<u}.
\end{align}
This in particular verifies that $ \mu_{*,u} $ solves the integral equation~\eqref{eq:inteq:}.

\subsection{B.2) Solving the variational problem~\eqref{eq:inteq}.}
Having solved \eqref{eq:inteq:} for generic $ u $, we now return to the variational problem \eqref{eq:inteq} (or consequently \eqref{eq:var}).
Consider $ (u_0,v_0) $ for which $ \frac{u_0}{2} - (Av_0^{1/2})/\pi = 0 $ and $ u_0+v_0=a_0 $,
or more explicitly
\begin{align}
	\label{eq:u0}
	u_0 = \frac{2A^2}{\pi^2}\Big( \sqrt{1+\frac{\pi^2a_0}{A^2}} -1 \Big),
	\quad
	v_0 = a_0-u_0= \frac{A^2}{\pi^2}\Big( \sqrt{1+\frac{\pi^2a_0}{A^2}} -1 \Big)^2.
\end{align}
For such $ (u_0,v_0) $, it follows from~\eqref{eq:mu*u}
that $ \mu_{*,u_0} $ gives a density (i.e.\ $ \mu_{*,u}(a) \geq 0 $) that satisfies the zero mass condition
\begin{align}
	\label{eq:0mass}
	\int_{\R} \mathrm{d} a\, (\mu_{*,u_0}(a)-\aden(a)) = 0.
\end{align}
Furthermore $ \mu_{*,u_0} $ solves the variational problem~\eqref{eq:inteq}.
From~\eqref{faeq}, we have
\begin{align*}
	-A\ind_\set{a<a_0} - 2\int_{\R} \mathrm{d} a' \, \frac{\mu_{*}(a')-\mu_\text{Airy}(a')}{a-a'}
	=
	g(a) \ind_\set{a<u_*}.
\end{align*}
Setting $ (u,v)=(u_0,v_0) $ in~\eqref{eq:g}
gives $ g(a)= (- |a-u_0|^{\frac12} - \tfrac{A}{\pi}\tan^{-1}(\sqrt{|a-u_0|/v_0}))\ind_\set{a<u_0} \leq 0 $.
This verifies that $ \mu_{*,u_0}(a) $ satisfies \eqref{eq:inteq2}.
In fact, $ u_0 $ is the \emph{only} value for which $ \mu_{*,u}(a) \geq 0 $ and satisfies \eqref{eq:inteq2}.
Also, $ \mu_{*,u_0} $ being the solution of \eqref{eq:inteq:} for $ u=u_0 $,
indeed satisfies \eqref{eq:inteq1}.
We have thus obtained the unique minimizer of $ \Sigma_J $:
\begin{align*}
	\mu_{*}(a)
	:=
	\mu_{*,u_0}(a)
	=
	\Big(
		\frac{1}{\pi}(a-u)^{\frac12} + \frac{A}{2\pi^2} \log \Big|\frac{\sqrt{a-u_0}+\sqrt{v_0}}{\sqrt{a-u_0}-\sqrt{v_0}}\Big|
	\Big)
	\ind_\set{a>u_0}.
\end{align*}

\subsection{B.3) Finding the minimum of $ \Sigma_J $ and $ \Sigma $.}

Having obtained the minimizer $ \mu_{*}=\mu_{*,u_0} $ of $ \Sigma_J $, we evaluate the minimum $ \Sigma_J(\mu_*) $.
We do so by first evaluating $ \Sigma(\mu_{*,u}) $ for generic $ u $, and specializing to $ u=u_0 $ later.
Consider
$
	F(a) = - \int_{\R} \mathrm{d}a'\, (\mu_*(a')-\aden(a')) \log|a-a'| .
$
Given that $ \frac{\mathrm{d}~}{\mathrm{d}a}F(a)=f(a) $  from \eqref{faeq}, and given the explicit expression of $ f(a) $,
we integrate $ f(a) $ to get
\begin{align*}
	F(a) -F(a_0) = G(a) -\tfrac{A}{2}(a_0-a)_+,
	\quad
	G(a) :=\int_{u}^{a} \mathrm{d} a' g(a') = -\ind_\set{a<u}\int_{a}^{u} \mathrm{d} a' g(a').
\end{align*}
We now calculate $ \Sigma_J(\mu_{*,u}) $.
\begin{align*}
	\Sigma_J(\mu_{*,u})
	=& A \int_u^{a_0} \mathrm{d}a \, (a_0-a)\mu_{*,u}(a) + \int_{\R^2} \log|a_1-a_2| \prod_{i=1}^2 \mathrm{d}a_i (\mu_{*,u}(a_i)-\aden(a_i))
\\	
	=& A \int_u^{a_0} \mathrm{d}a \, (a_0-a)\mu_{*,u}(a) + \int_0^\infty \mathrm{d}a \, (\mu_{*,u}(a)-\aden(a)) F(a).
\end{align*}
Given \eqref{eq:0mass},
%For the density $ \mu_{*,u} $, from the definition~\eqref{eq:mu*u},
%it is straightforward to verify that $ \int_{0}^\infty \mathrm{d}a\,(\mu_{*,u}(a)-\aden(a)) =0 $.
%%{\red I: I am forgetting why this is true. please point to the reason}
%Using this,
in the last integral above we may replace $ F(a) $ with $ F(a)-F(a_0) $. Combine this with the fact that $\int_0^\infty \mathrm{d}a \mu_{*,u}(a)G(a)=0$, we arrive at
\begin{align}
	\label{eq:Sigma*}
	\Sigma_J(\rho_{*,u}) =
	\frac{A}{2} \int_u^{a_0} \mathrm{d}a \, (a_0-a)\mu_{*,u}(a)
	+ \frac{A}{2} \int_{0}^{a_0} \mathrm{d}a \, (a_0-a)\aden(a)
	-
	\int_{0}^{u} \mathrm{d}a\, G(a) \aden(a).
\end{align}
Set $ \aDen(a) = \int_0^a \mathrm{d}a'\,\aden(a') = \frac{2}{3\pi}a^{3/2}\ind_\set{a>0} $.
Applying integration by parts to the last term in~\eqref{eq:Sigma*} gives
\begin{align*}
	-
	\int_{0}^{u} \mathrm{d}a\, G(a) \aden(a)
	=
	\int_{0}^{u} \mathrm{d}a\, g(a) \aDen(a).
\end{align*}
Plugging this back into~\eqref{eq:Sigma*} yields
\begin{align*}
	\Sigma_J(\rho_{*,u}) =
	\frac{A}{2} \int_u^{a_0} \mathrm{d}a \, (a_0-a)\mu_{*,u}(a)
	+ \frac{A}{2} \int_{0}^{a_0} \mathrm{d}a \, (a_0-a)\aden(a)
	+ \int_{0}^{u} \mathrm{d}a\, g(a) \aDen(a).
\end{align*}
This integral is evaluated by Mathematica, giving
\begin{align*}
	\Sigma_J(\rho_{*,u})
	=
	\Sigma_J(u,a_0-u),
	\quad
	\text{where }
	\Sigma_J(u,v)
	:=
	\tfrac{1}{12}u^3
	+\tfrac{2}{3\pi}Auv^{\frac32}
	-\tfrac{1}{2\pi^2} A^2v^{2}
	+ \tfrac{4}{15\pi}Av^{\frac52}.
\end{align*}
%
%Next, we optimize $ \Sigma_J(u,a_0-u) $ over $ u $. Differentiating in $ u $ gives
%$
%	\partial_u \Sigma_J(u,a_0-u)
%	=
%	(\tfrac{u}{2}-\tfrac{Av^{\frac12}}{\pi})^2|_{v=a_0-u} \geq 0,
%$
%so $ \Sigma_J(u,a_0-u) $ strictly increases in $ u $.
%However, since we require that $\mu_{*,u}$ is a probability density, we must have $ \mu_{*,u}(a) \geq 0 $. This implies that $ \frac{u}{2}-\frac{Av^{\frac12}}{\pi} \geq 0 $.
%These constraints imply that the minimum is achieved at the boundary point $ u_0 $ which solves $ u_0+v_0=a_0 $ and $ \frac{1}{2}u_0-\frac{1}{\pi}Av_0^{1/2}=0 $.
%Solving this gives
%
Now, specializing at $ u=u_0 $ gives
\begin{align*}
%	\mu_*(a) &= \Big( \frac{1}{\pi}(a-u_0)^{\frac12} + \frac{A}{2\pi^2} \log\Big|\frac{\sqrt{a-u_0}+\frac{\pi u_0}{2A}}{\sqrt{a-u_0}-\frac{\pi u_0}{2A}}\Big| \Big)\mathbf{1}_{\{x>u_0\}},
%\\
	\Sigma_J(\mu_*) = \Sigma(u_0,a_0-u_0)
%	&=
%	- \frac{4A^6}{15\pi^6} - \frac{2A^4a_0}{3\pi^4} - \frac{A^2a_0^2}{2\pi^2}
%	+ \frac{4A^6}{15\pi^6}\Big( 1+\frac{\pi^2a_0}{A^2} \Big)^{\frac52}
%\\
	&=
	A^6
	\Big(
		- \frac{4}{15\pi^6} - \frac{2a}{3\pi^4} - \frac{a^2}{2\pi^2}
		+ \frac{4}{15\pi^6}\big( 1+\pi^2a \big)^{\frac52}
	\Big)\Big|_{a=a_0A^{-2}}.
\end{align*}

This gives the minimum of $ \Sigma_J $,
and we now return to $ \Sigma $.
Recall that $ I_\text{Airy}(\mu) = J_\text{Airy}(\mu)+U(\mu) $,
where $ U(\mu)$ is a nonnegative functional.
Consequently, $ \Sigma(\mu) \geq \Sigma_J(\mu) $, and hence
\begin{align*}
	\min_{\mu} \Sigma(\mu) \geq \min_{\mu} \Sigma_J(\mu) = \Sigma_J(\mu_*).
\end{align*}
Further, since $ \mu_*(a) $ vanishes for $ a<0 $ (since $ u_0>0 $),
we have $ U(\mu_*)=0 $ so that $ \Sigma(\mu_*)=\Sigma_J(\mu) $.
Thus, the minimizer and minimum we solved for with respect to $ \Sigma_J $ in fact also apply to $ \Sigma $. Since $ \Phi_{-}(z)=\min_{\mu} \Sigma(\mu) $, this confirms the formula in \eqref{eq:Phi} and the calculation of \cite{sasorov2017large}.

To compare our result to those of~\cite{dean} on forced Coulomb-gas, consider $ A\to\infty $.
This corresponds to an infinite potential wall at $ a=a_0 $  forcing all eigenvalues into region $ a>a_0 $.
As $ A\to\infty $ we have $ u_0\to a_0 $, and
\begin{align*}
	\mu_*(a) \longrightarrow \big(\tfrac{1}{\pi}(a-a_0)^{1/2} + \tfrac{a_0}{2\pi}(a-a_0)^{-\frac12}\big)\ind_\set{a>a_0},
	\quad
	\quad
	\Sigma(\mu_*) \longrightarrow \tfrac{1}{12} a_0^3.
\end{align*}
This recovers the edge limit of the results in \cite{dean}.\\

It is useful to summarize and display the solution of the following variational problem
for arbitrary constants $A,B$ as
\be
\min_\mu \left[
	 A \int_{\mathbb{R}}\mathrm{d}a \, \mu(a)(-z-a)_{+} + B \,  I_\text{Airy}(\mu) \right] = \frac{A^6}{B^5} \Phi_-(z \frac{B^2}{A^2})
\ee
where $\Phi_-(z)$ is given in Eq. \eqref{eq:Phi}. This readily applies to the half-space
problem as mentioned in the text with the choices $A=B=\frac{1}{2}$ leading
to $\Phi^{\text{half-space}}_{-}(z) = \frac{1}{2} \Phi_{-}(z)$. It also applies to
the partition sum, defined in Eq. (1.12) in the arXiv version of Ref. \cite{VadimSodin18}, with $\alpha=t^{1/3}$,
of a directed polymer of length $2 t$ in a static Brownian random potential of amplitude
$\frac{1}{\sqrt{\beta}}$, plus a linear potential in a half space. Let us call $e^{{\cal H}_\beta(t)}$ precisely
that formula (1.12) there. In that work ${\cal H}_\beta(t)$ is shown to have the same distribution
as ${\cal H}(t)$ defined here for our full space problem for $\beta=2$, and
for our half-space problem for $\beta=1$. For general $\beta$ and $u>0$
\be
\overline{ \exp\big( - \tfrac{\beta^2}{4} u \, e^{{\cal H}_\beta(t)}\big)  }
= \mathbb{E}_{\beta} \left[ \prod_{i=1}^{+\infty} \frac{1}{(1 + u \, e^{t^{1/3} \mathbf{a}_i})^{\beta/2}} \right]
\ee
where $\mathbb{E}_{\beta}[..]$ denotes the expectation over the Airy$_\beta$ point process.
Setting $u=e^{s t^{1/3}}= e^{-z t}$, the above result with $A=B=\frac{\beta}{2}$ implies that the
large deviation crossover rate function associated to ${\cal H}_\beta(t)$ is
$\Phi^{\beta}_{-}(z) = \frac{\beta}{2} \Phi_{-}(z)$ as announced in the text. \\

Finally, consider the full space Brownian IC (with possible drifts)
$h(x,0)=B(x) - w |x|$, where $B(x)$ is a standard Brownian motion.
It is convenient to scale $w = t^{-1/3} \tilde w$ at fixed $\tilde w$. The case $\tilde w=0$ corresponds
to the stationary IC. Let us list the arguments which support the conjecture that
$\Phi^{\rm Br}_{-}(z)=\Phi^{\rm droplet}_{-}(z)$ for any $\tilde w$.

\begin{itemize}

\item The left tail of the Baik-Rains distribution, which is the late time distribution of the height field associated to the stationary IC, suggests that upon matching (in our units)
$\Phi^{\rm Br}_{-}(z) \simeq_{z \to 0^-} \frac{1}{12} |z|^3
\simeq \Phi^{\rm drop}_{-}(z)$.

\item It is shown in \cite{KrajLedou2018,KrajLeDouProlhac18} from the calculation of the first cumulant
that $\Phi^{\rm Br}_{-}(z) \simeq_{z \to - \infty} \frac{4}{15 \pi} (-z)^{5/2}
\simeq \Phi^{\rm drop}_{-}(z)$. Furthermore explicit calculation of the next (second) cumulant
shows that the next term in the large negative $z$ expansion of
$\Phi^{\rm Br}_{-}(z)$  is $- \frac{z^2}{2 \pi^2}$, identical to the one of $\Phi^{\rm drop}_{-}(z)$.

\item The so called Baik-Rains kernel associated to the Brownian IC at large time is a finite rank modification
of the Airy kernel and relates to a finite rank perturbation of the same Coulomb gas we are considering. This finite rank perturbation contributes terms  proportional to $N$ which are subdominant compared
to $N^2$ and hence should not modify the present calculation.

\end{itemize}

\section{C) Details in deriving Eqs. \eqref{eq:1}}
We describe the main ideas which yield \eqref{eq:1} (in fact, a complete mathematically rigorous proof is given in \cite{CorwinGhosalTail}). The analysis starts by reducing the tail bounds in \eqref{eq:1} to corresponding bounds on the r.h.s of \eqref{id}.
%\be
%\mathbb{E}_{\Airy}\left[\exp\left(-\sum_{i=1}^\infty \varphi_{t,s}(\mathbf{a}_i )\right)\right].\label{eq:eAiry}
%\ee
In order to bound this expectation it is useful to understand the typical locations of the $\mathbf{a}_i$. At a rough level, this can be deduced from the density $\rho(a)$. More precisely we may use the fact that the Airy PP coincides with the spectrum of the `stochastic Airy operator' \cite{StochasticAiry}. Define the `Airy operator' $\mathcal{A}$ by  $(\mathcal{A}f)(x) = -f^{\prime\prime}(x) +xf(x)$ and the stochastic Airy operator $\mathcal{H}_{\beta}$ with inverse temperature $\beta>0$ by $(\mathcal{H}_\beta f )(x) = (\mathcal{A}f)(x) + \frac{2}{\sqrt{\beta}}f(x)B^{\prime}(x)$ where $B(x)$ is a Brownian motion. A function $f$ is an eigenfunction for $\mathcal{H}_{\beta}$ (resp.\ $\mathcal{A}$) if $\int_0^{\infty}\big(f^{\prime}(x)^2+(1+x)f(x)^2\big)\mathrm{d}x<\infty$, $f(0)=0$ and $(\mathcal{H}_{\beta}f)(x) = \Lambda f(x)$ (resp.\ $(\mathcal{A}f)(x) = \lambda f(x)$). The ordered eigenvalues $\lambda_1<\lambda_2<\cdots$ of $\mathcal{A}$ are such that $\mathrm{Ai}(x-\lambda) =0$ exactly that $\lambda=\lambda_i$. Classical estimates show that $\lambda_n\approx \big(\frac{3\pi}{2}n\big)^{2/3}$. Ref. \cite{StochasticAiry} proves that if $\Lambda_1\leq \Lambda_2<\cdots$ are the eigenvalues of $\mathcal{H}_{\beta}$, then for $\beta=2$, in distribution $\mathbf{a}_i = -\Lambda_i$ (recall the $\mathbf{a}$ represent the APP). Likewise, the GOE and GSE version of the Airy PP coincide with the spectrum of $\mathcal{H}_{\beta}$ for $\beta=1$ and $4$.

Since formally, $\mathcal{H}_{\beta}$ converges to $\mathcal{A}$ as $\beta\to \infty$, it is natural to hope that the (random) spectrum of $\mathcal{H}_{\beta}$ is probabilistically close to the (deterministic) spectrum of $\mathcal{A}$. This is substantiated through the following result:
\begin{lemma}
For any $\beta>0$, define the random variable $C^\beta_{\varepsilon}$ as the minimal value of $C$ such that for all $k\geq 1$, $(1-\varepsilon)\lambda_k -C\leq \Lambda_k^\beta \leq (1+\varepsilon)\lambda_k + C$. Then, for any $\delta>0$ there exists $s_0, \kappa>0$ such that for all $s\geq s_0$,
\be
\mathbb{P}(C^{\beta}_{\varepsilon}\geq \beta^{-1/2} s) \leq \kappa \exp(-\kappa s^{1-\delta}).\label{eq:Ceps}
\ee
\end{lemma}
This result shows that up to a controllable error, the Airy PP $\mathbf{a}_n$ are uniformly (in a multiplicative sense) close to their typical locations $-\big(\frac{3\pi}{2}n\big)^{2/3}$. Simply plugging these typical locations into the r.h.s of \eqref{id} yields the $\frac{4}{15\pi}s^{5/2}$ tail behavior. The other terms (namely the cubic tail behavior) in \eqref{eq:1} come from the effect of deviations of the Airy PP from these locations. Alone, the bound in \eqref{eq:Ceps} is not sufficient to estimate this effect.

We need to develop new precise and uniform estimates on the deviations of the Airy PP on large intervals. Define the counting function $\chi$ for the Airy PP so that for any interval $B$, $\chi(B)=\#\big\{i:\mathbf{a}_i\in B\big\}$. Define intervals $B_0=[-s,\infty)$ and $B_k=[-ks,-(k-1)s)$ for $k\geq 1$. Using the one and two-point correlation functions of the APP, \cite{Soshnikov} shows that for any $s>0$, up to bounded errors as $s\to \infty$, $\mathbb{E}\big[\chi(B_0)\big]\approx \frac{2}{3\pi} s^{3/2}$,   $\mathrm{var}\big(\chi(B_k)\big)\approx \frac{11}{12\pi^2} \log(s)$ for all $k\geq 0$. From this follows:
\begin{lemma}
For any $k\geq 0$, there exists $s_0$ such that for all $s\geq s_0$ and $c>0$,
\be
\mathbb{P}\Big(\chi(B_k) - \mathbb{E}\big[\chi(B_k)\big]\geq cs^{3/2}\Big) \leq \exp\big(-\frac{3}{2}cs^{3/2}\big).\label{eq:thm15}
\ee
\end{lemma}
This estimate follows from the fact \cite[Section 4.2]{AGZ} that for any compact set $B$, $\chi(B)$ equals (in distribution) the sum of independent Bernoulli ($0$ or $1$ valued) random variables with parameters given by the eigenvalues of $\mathbf{1}_{B}K_{\mathbf{Ai}}\mathbf{1}_{B}$, and an application of Bennett's inequality \cite{Bennett} which states that for independent Bernoulli random variables $X_1,\ldots, X_n$ setting $S=X_1+\cdots+X_n$ and $\sigma^2=\mathrm{var}(S)$, then $\mathbb{P}(S>t) \leq \exp\big(-\sigma^2 h(t/\sigma^2)\big)$ where $h=(1+u)\log(1+u)-u$. Using \eqref{eq:thm15}, we are able to establish the second line of \eqref{eq:1}. \eqref{eq:thm15} provides an upper bound on the number of $\mathbf{a}_i$ in various intervals which translates into the necessary lower bound on the expectation on the r.h.s of  \eqref{id}.% (which, as explained earlier, yields ~\eqref{eq:2}).

In order to establish the first line of \eqref{eq:1} we must estimate an upper bound on the expectation in the r.h.s of \eqref{id}. This is done by establishing a lower bound on the number of $\mathbf{a}_i$ exceeding given (large negative) values. In particular:
\begin{lemma}
For any $\delta>0$ there exists $s_0>0$ such that for all $s\geq s_0$ and $c>0$,
\be
\mathbb{P}(A) \leq \exp\big(-cs^{3-\delta}\big).\label{eq:thm14}
\ee
where $A:=\Big\{\chi([-s,\infty)) - \mathbb{E}\big[\chi([-s,\infty))\big]\leq -cs^{3/2}\Big\}$.
\end{lemma}
This result is considerably harder to prove than \eqref{eq:thm15}. Markov's inequality shows that for any $\lambda>0$,
$$
\mathbb{P}(A) \leq \exp\Big(-\lambda c s^{3/2} +\lambda \mathbb{E}\big[\chi([-s,\infty))\big]\Big)\, F(-s;\lambda)
$$
where the cumulant generating function
$F(-s;\lambda):= \mathbb{E}\Big[\exp\big(\lambda \chi([-s,\infty))\big)\Big].$
By taking $\lambda = s^{3/2 - \delta}$ and using our earlier estimate on $\mathbb{E}\big[\chi([-s,\infty))\big]$, proving \eqref{eq:thm14} reduces to showing:
\begin{lemma}
For all $\delta>0$, there exists $s_0,\kappa>0$ such that for all $s\geq s_0$
\be
F(-s;s^{3/2-\delta}) \leq \exp\big(-\kappa s^{3-\delta}\big).  \label{eq:Fest}
\ee
\end{lemma}
To prove \eqref{eq:Fest} we appeal to a connection between $F(-s;\lambda)$ and the Ablowitz-Segur (AS) solution to the Painlev\'{e} II equation \cite{AS70s,Bothner15}. For $\gamma=1-e^{-v}$,
\begin{align}
F(x;v) = \exp\Big(-\int_{x}^{\infty} (y-x) u^2_{\mathrm{AS}}(y;\gamma) \mathrm{d}y\Big),\label{eq:Fas}
\end{align}
where $u_\mathrm{AS}(y;\gamma)$ solves the Painlev\'{e} II equation
$
u_\mathrm{AS}^{\prime\prime}(x;\gamma) = x u_\mathrm{AS}(x;\gamma) + 2 u_\mathrm{AS}^3(x;\gamma)
$
with AS boundary condition
$
u_\mathrm{AS}(x;\gamma) \sim_{x\to \infty} \sqrt{\gamma} \frac{x^{-1/4}}{2\sqrt{\pi}} e^{-\frac{2}{3}x^{3/2}}\, \big(1+o(1)\big)$.

The AS solution has received attention recently in \cite{BdCP09,Bothner15,BB17} since $\gamma K_{\mathbf{Ai}}$ is the correlation kernel for a `thinned' version of the Airy PP (where each particle is removed with probability $1-\gamma$). Our bound on $F(x;v)$ also provides a tail bound on that process. In order to establish \eqref{eq:Fest}, we utilize an explicit formula for the behavior of $u_\mathrm{AS}(-s;\gamma)$ with $\gamma=1-\exp(-s^{3/2-\delta})$ as $s\to \infty$ (and $\delta>0$ arbitrary), computed in \cite{Bothner15} (via a $2\times 2$ Riemann-Hilbert problem steepest descent analysis). The asymptotic form given in \cite{Bothner15} involves Jacobi elliptic and theta functions, and is highly oscillatory. The proof of \eqref{eq:Fest} requires controlling these oscillations in order to estimate the integral in \eqref{eq:Fas}.

\end{widetext}
\end{document}